\documentclass[iop]{emulateapj}

\usepackage{amssymb}
\usepackage[multiple]{footmisc}
\usepackage{textcomp}
\usepackage{color}
\usepackage{graphics}
\usepackage{xcolor}
\usepackage{natbib}
\usepackage{appendix}
\usepackage{amsmath}
\usepackage{vector}

\shorttitle{Photo-zs of LRGs from the SDSS DR7}
\shortauthors{N. Greisel et al.}

\newcommand{\snmad}{\sigma_{\mathrm{NMAD}}}
\newcommand{\zspec}{z_{\mathrm{spec}}}
\newcommand{\zphot}{z_{\mathrm{phot}}}
\newcommand{\dz}{\Delta z}
\newcommand{\photoz}{\texttt{PhotoZ}}
\newcommand{\sedfit}{\texttt{SEDfit}}

\newcommand{\br}{4000\,\AA{} break}
\newcommand{\bpz}{\texttt{BPZ}}
\newcommand{\lephare}{\texttt{LePhare}}
\newcommand{\ezgal}{\texttt{EzGal}}

\definecolor{bluet}{HTML}{4169e1}
\definecolor{greent}{HTML}{20b2aa}
\definecolor{redt}{HTML}{b22222}
\definecolor{magenta}{HTML}{ee00ee}

\begin{document}

\title{PHOTOMETRIC REDSHIFTS AND SYSTEMATIC VARIATIONS IN THE SPECTRAL ENERGY DISTRIBUTIONS OF LUMINOUS RED GALAXIES FROM THE SDSS DR7}

\author{N. Greisel\altaffilmark{1,2}, S. Seitz\altaffilmark{1,2}, N. Drory\altaffilmark{2,3}, R. Bender\altaffilmark{1,2}, R.~P. Saglia\altaffilmark{1,2}, J. Snigula\altaffilmark{1,2}}
\email{greisel@usm.lmu.de}
\altaffiltext{1}{Universit\"{a}ts-Sternwarte M\"{u}nchen, Ludwig-Maximilians-Uni\-ver\-si\-t\"{a}t M\"{u}nchen, Scheinerstr. 1, D-81679 M\"{u}nchen, Germany}
\altaffiltext{2}{Max-Planck-Institut f\"{u}r extraterrestrische Physik, Giessenbachstra\ss e, D-85748 Garching, Germany}
\altaffiltext{3}{Instituto de Astronom\'ia, Universidad Nacional Aut\'onoma de M\'exico, Avenida Universidad 3000, Ciudad Universitaria, C.P. 04510, D.F., Mexico}

\begin{abstract}
We describe the construction of a template set of spectral energy distributions (SEDs) for the estimation of photometric redshifts of luminous red galaxies (LRGs) with a Bayesian template fitting method.
By examining the color properties of several publicly available SED sets within a redshift range of $0<z\lesssim0.5$ and comparing them to Sloan Digital Sky Survey (SDSS) Data Release 7 data,
we show that only some of the investigated SEDs approximately match the colors of the LRG data throughout the redshift range, however not at the quantitative level required for precise photometric redshifts.
This is because the SEDs of galaxies evolve with time (and redshift) and because at fixed redshift the LRG colors have an intrinsic spread such that they cannot be matched by one SED only.
We generate new SEDs by superposing model SEDs of composite stellar populations with a burst model, allowing both components to be reddened by dust, in order to match the data in five different redshift bins.
We select a set of SEDs which represents the LRG data in color space within five redshift bins, thus defining our new SED template set for photometric redshift estimates.
The results we get with the new template set and our Bayesian template fitting photometric redshift code (\photoz) are nearly unbiased, with a scatter of $\sigma_{\Delta z}=0.027$ (including outliers), a fraction of catastrophic outliers ($|\zphot-\zspec|/(1+\zspec)>0.15$) of $\eta=0.12\%$, and a normalized median absolute rest frame deviation (NMAD) of $\snmad=1.48\times\mathrm{MAD}=0.017$ for non-outliers.
We show that templates that optimally describe the brightest galaxies ($-24.5\leq M_R\leq-22.7$) indeed vary from $z=0.1$ to $z=0.5$, consistent with aging of the stellar population.
Furthermore, we find that templates that optimally describe galaxies at $z<0.1$ strongly differ as a function of the absolute magnitude of the galaxies, indicating an increase in star formation activity for less luminous galaxies.
Our findings based on the photometry of the SDSS LRGs and our SED template fitting are supported by comparison to the average SDSS LRG spectra in different luminosity and redshift bins.
\end{abstract}


\section{Introduction} 
The primary goal of our work is to generate new spectral energy distributions (SEDs) for luminous red galaxies (LRGs) to obtain precise photometric redshift measurements from template fitting methods.
For the analysis of large scale structure (LSS) based on photometric redshift distances it is important to precisely know the photometric redshift errors and biases and to make these as small as possible.
For dark energy constraints based on shear tomography it is most important to know the errors and biases precisely \citep{2006ApJ...636...21M}, but on the other hand the rms-size of the photometric redshift errors determines how many tomographic redshift bins can be analyzed for a given survey.
In addition, if the bias changes strongly as a function of the (true) redshift, then it cannot be treated as being constant within a given tomographic interval anymore, which makes the analysis of such data more difficult.
Another reason why small photometric redshift uncertainties are advantageous is the following.
When the baryonic acoustic oscillation (BAO) scale is extracted from galaxy clustering measurements obtained with distances based on photometric redshifts (see \citealt{2011MNRAS.411..277S}, and \citealt{2009ApJ...691..241B}  for the case of the transversal BAO scale, and \citealt{2009JCAP...04..008R} for the case of the radial direction), the quality of the result dramatically depends on the photometric redshift errors (see Figure~3 of \citealt{2009ApJ...691..241B}, and Figure~1 of \citealt{2011MNRAS.411..277S}).
When galaxy properties (colors, star formation rate, dark matter halo properties) are studied as a function of their environment density based on photometric redshift distances (see, e.g., \citealt{2007ApJS..172..284C} and \citealt{2011MNRAS.413.1678C} for an example), redshift errors introduce biases in density, because galaxies are being preferentially scattered out of high density environments into the low density environments.
In these cases it is also very important to have a low SED dependence of bias, or to at least know the bias very well as a function of SED and redshift.
Finally, redshift errors imply large errors of the estimated absolute luminosities and stellar masses (see \citealt{2012MNRAS.420.3240N}, Figure~13).
If one then studies average galaxy properties as a function of absolute luminosity or stellar mass, one has to account for contamination from galaxies originating from other luminosity bins.\\
In summary, it is important to know bias and redshift scatter, but it is also important to aim for a scatter that is as small as possible, since any scatter corresponds to a loss of information.\\
We will generate new template SEDs which optimally reproduce the LRG colors by fitting composite stellar populations (CSPs) and an extincted starburst component to the broadband photometric Sloan Digital Sky Survey (SDSS) data.
We thus also obtain hints for the mechanism causing the observed changes, i.e., a luminosity and redshift dependence of star formation history (SFH).
Broadband photometry however is not the best method to study this, given that different ages and metallicities can produce fairly similar colors.
This degeneracy can only be rigorously broken with spectroscopic information.
We could use priors (i.e., exclude low metallicities) from the beginning, and then interpret the best fitting SEDs in terms of SFH.
Instead, we want to make use of the full variety of different colors that can be generated even with implausible ingredients (i.e., age\textendash metallicity combination, but also extinction which changes the slope of an SED) to increase the flexibility of SEDs that can be generated.
We will see in Section~\ref{sec:UV}, however, that the broadband near-UV (NUV) and far-UV (FUV) colors of our newly generated SEDs match the sequence of colors of quiescent to star-forming early-type galaxies (ETGs) classified by spectroscopic analyses.
This means that generating SEDs by fitting galaxy colors with a composite population plus a burst population is not just a method to increase the freedom of SEDs in the $u$- to $z$-band (rest frame) wavelength ranges, but that it is very plausible that some of the LRGs do show signs of recent star formation.
In this paper we concentrate on the estimation of photometric redshifts (photo-$z$s) for spectroscopically observed LRGs
from the SDSS Data Release 7 \citep{2000AJ....120.1579Y,2009ApJS..182..543A}.
LRGs mostly consist of ETGs and show rather uniform SEDs \citep{1983ApJ...264..337S,2003ApJ...585..694E}.
Most importantly, their SEDs have a pronounced break at 4000\,\AA\ which is due to an accumulation of metal lines.
The \br\ together with the spectral uniformity are the reasons why photometric redshift estimations are very precise for red galaxy types.
LRGs are also of great cosmological interest: They are among the most luminous galaxies existing, are strongly correlated with each other, since they are preferentially located in dense regions of the universe, they closely map its LSS.\\
This paper is organized as follows: In Section~\ref{sec:PhotoZ} we briefly present the photometric redshift code used in this paper.
After that we will shortly explain the data set and investigate the differences between the colors of the data and specific model SEDs (Section~\ref{sec:datamod}).
In Section~\ref{sec:sedfitmodels} we present the method with which we generate new SEDs based on the LRG photometry.
The photometric redshift results using the new template SEDs are analyzed in Section~\ref{sec:results} and compared to public SDSS photo-$z$s in Section~\ref{sec:comparison}.
A discussion about the properties of the new SEDs is given in Section~\ref{sec:discussion}, where we investigate the FUV and NUV colors in Section~\ref{sec:UV} and show systematic variations in the SEDs in Section~\ref{sec:zvsmabs}.
We show the results of SED fitting different model sets to the LRG photometry in Appendix~\ref{app:models}.
The colors of the created model SEDs will be investigated in greater detail in Appendix~\ref{app:colcol}.\\
Throughout this paper we assume a cosmology with $\Omega_m=0.3$, $\Omega_\Lambda=0.7$ and $H_0=70\,\mathrm{km\,s}^{-1}\mathrm{Mpc}^{-1}$, while magnitudes are given in the AB system.

\section{Photometric Redshifts with Template Fitting Methods and the \photoz\ Code}
\label{sec:PhotoZ}
The techniques to estimate photometric redshifts can be divided into two main categories: empirical and template fitting methods.
The former use a training set of galaxies with known spectroscopic redshifts and photometry to derive an empirical relation between the redshift and the photometric observables.
The latter fit template SEDs to the photometry of objects, thereby determining the most probable redshift \citep{1999MNRAS.310..540A,2000ApJ...536..571B,2000A&A...363..476B,2000AJ....119...69C,2001defi.conf...96B,2002A&A...386..446L,2004MNRAS.353..654B,2006MNRAS.372..565F,2008ApJ...686.1503B}.
The simplest approach is to perform solely a $\chi^2$ minimization of the difference between the expected colors of a SED template and observed photometry, whereas hybrid photometric redshift methods combine a maximum likelihood approach together with an optimization routine to enhance the initial template set (e.g., by using a spectroscopic training set).
Bayesian methods, on the other hand, use $\chi^2$ minimization, but additionally provide the opportunity to introduce redshift and luminosity priors which assign probability functions to the templates, thus enabling one to lift possible degeneracies where models are equally probable at different redshifts.
The model SEDs for template matching methods can either be derived empirically \citep[e.g.,][]{1980ApJS...43..393C}, from stellar population models \citep[e.g.,][]{2003MNRAS.344.1000B}, or a combination of both.
The advantage of template fitting is that it does not only provide photometric redshift estimates, but also rest frame properties of the best fit models such as the SED type and the absolute magnitude.
Second, if good SEDs are available, they can be used with any data set (independent of its photometric depth, or the filter system).
Since several SED types and redshifts can match the colors of an observed object equally well, it is very important to optimize the redshift and the luminosity priors for each type, depending on the filter set and the data, i.e., photometric depth and galaxy properties (which, in turn, depend on the galaxy selection function).

\subsection{The \photoz\ Code}
In this work we use the Bayesian template fitting photometric redshift code \photoz\ \citep{2001defi.conf...96B}.
The calculation of the probabilities of the model-redshift combinations is based on Bayes' theorem:
\begin{eqnarray}
P(\mathbf{\uvec{\mu}}|C,m)\propto \mathcal{L}(C,m|\mathbf{\uvec{\mu}})\cdot P(\mathbf{\uvec{\mu}}),
\end{eqnarray}
where $C$ and $m$ denote the colors and magnitudes of the photometric data, and $\mathbf{\uvec{\mu}}$ are the model parameters $z$ and $M$.
$\mathcal{L}(C,m|\mathbf{\uvec{\mu}})$ is the likelihood function which is proportional to $\exp\left(-\chi^2/2\right)$.
The second factor is the prior distribution $P(\mathbf{\uvec{\mu}})=P(M|T)\cdot P(z|T)$, with the probability functions of absolute magnitude $M$ and redshift $z$ for a template $T$.
These priors are specified a priori and enable us to assign specific probability distributions in redshift and absolute magnitude for every SED template.
Both are parameterized as proportional to $P(x|T)=\exp\left(-\ln(2)\left(\frac{x-\hat{x}}{\sigma}\right)^p\right)$, where $\hat{x}$ is the most probable value of $x$, $x=M$ or $x=z$.
The priors were chosen in this functional form to be able to vary the transitions between regions of different probabilities in such a way that they may be both smooth and have a box-like shape.
In this formula the integer $p$ controls the steepness of the probability decrease left and right of $\hat{x}$: The larger $p$, the steeper the decrease (odd $p$ values make no sense, of course, as the integral over the probability would diverge).
We set the $z$ prior to have a Gaussian shape ($p=2$), in order to have a shallow transition from high to low probabilities.
We will make particular use of the $z$ prior  later on when using SEDs specifically designed for different redshift ranges.
For the luminosity prior we set $\hat{x}=\hat{M}_R=-21$, $\sigma=3$, and $p=6$, as luminosities $M_R\notin[-24,-18]$ are very rare among LRGs (see also Section~\ref{sec:zvsmabs}, Figure~\ref{fig:ZvsMabs_int3_zvsM}) and are thereby excluded.
Within this interval however, the luminosity probabilities are virtually equally distributed.\\
Although the LRG sample contains only objects with $z\lesssim0.5$, the allowed photometric redshift range for all \photoz\ runs is $[0,5.0]$.
In this way we can assure that the photometric redshift results truly are below $z\approx0.5$, although a larger redshift range is allowed.
This is important for future work with deeper surveys where the redshift range is not known a priori.\\
The \photoz\ code has been extensively tested and was applied to a number of photometric catalogs \citep{2001MNRAS.325..550D,2004A&A...421...41G,2005ApJ...633L...9F,2008arXiv0811.3211B,2008MNRAS.383.1319G,2013arXiv1303.6287B}.
It has also been successfully applied to the Pan-STARRS1 medium deep fields and is part of the PS1 Photometric Classification Server \citep{2012ApJ...746..128S}.

\section{The Data and Model SEDs}
\label{sec:datamod}

\subsection{The LRG Sample}
The data in this paper contains objects from the spectroscopic set of the SDSS Data Release 7 LRG sample \citep{2001AJ....122.2267E}.
The selection criteria for LRGs in SDSS separate them from the other galaxy types by cuts in color space, taking their evolution with redshift into account.
Objects passing these criteria are flagged accordingly and can be obtained via the Catalog Archive Server\footnote{\url{http://cas.sdss.org/astrodr7/en/}} (CAS).
SDSS LRGs are color-selected in such a way that their SEDs have prominent \br s.
We remove objects from the SDSS spectroscopic LRG sample that do not pass the color cuts developed by \citet{2005MNRAS.359..237P}, which differ marginally from the selection criteria of \citet{2001AJ....122.2267E}.
For a reasonable evaluation of the photometric redshift quality later on, we require that the spectroscopic redshift errors fulfill $\dz_{\mathrm{spec}}\leq0.002$ (the redshift resolution of the \photoz\ runs), and that the spectroscopic redshifts were determined with high confidence ($>99\%$).
The studied LRG sample contains approximately $140{,}000$ objects.\\
Unless mentioned otherwise, $ugriz$ refer to the extinction corrected \citep{1998ApJ...500..525S} model magnitudes of SDSS in the AB system \citep{1983ApJ...266..713O}.

\subsection{Colors of SED Templates versus Colors of SDSS LRGs}
\label{sec:modelcolors}
In order to obtain decent photometric redshift estimates with template fitting methods, the model-set has to represent the data well in all available colors.
In this section we compare the color\textendash redshift relations for several red SEDs with the observed SDSS LRG colors.
The red SEDs are taken from model sets provided in commonly used template fitting photometric redshift codes.
Additionally, we examine other SEDs from stellar population modeling such as those of \citet[][hereafter BC03]{2003MNRAS.344.1000B} and \citet[][hereafter M09]{2009MNRAS.394L.107M}.

\subsubsection{The \photoz\ Standard Template Set}
\begin{figure}
\includegraphics[width=0.5\textwidth]{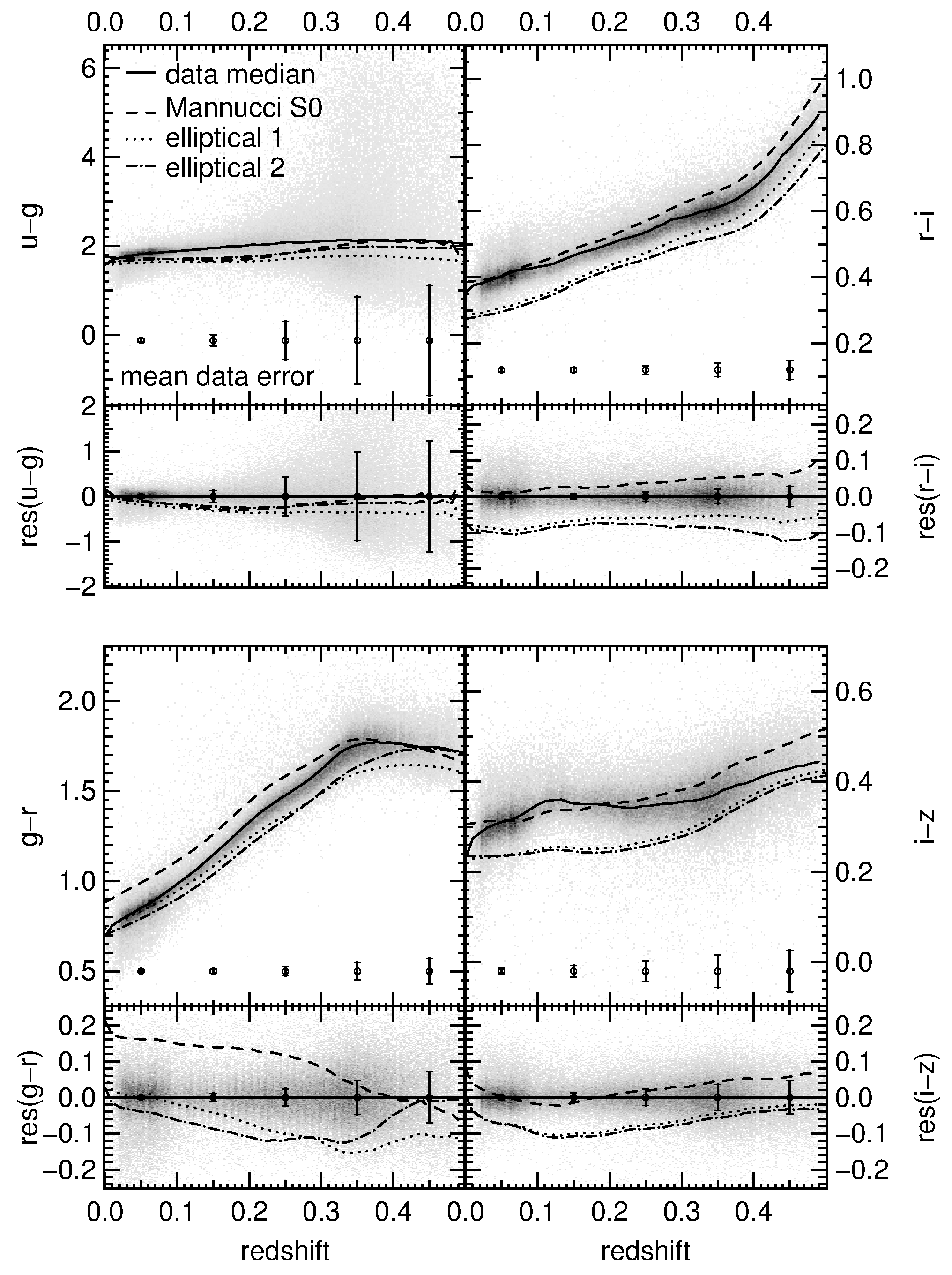}
\caption{
Upper panels: redshift vs. colors for the three red model SEDs that were previously used with the \photoz\ code.
	  Lower panels: color residuals between models and data.
	  LRGs are displayed in gray and their median is shown by a solid line.
	  Error bars show mean data errors in five redshift bins.
	  The dashed line indicates the Mannucci S0 SED, whereas the colors of the two other red SEDs created for the HDF are shown with dotted and dash-dotted lines.
	  }
\label{fig:ZvsCOL_std}
\end{figure}
First, we investigate the three red template SEDs that have been used with the \photoz\ code until now \citep{2004A&A...421...41G,2005ApJ...633L...9F,2008MNRAS.383.1319G,2008arXiv0811.3211B}.
One of the models originates from \citet{2001MNRAS.326..745M} and describes a galaxy with type S0.
The other two SEDs were generated synthetically from single stellar populations of \citet{1993ApJ...405..538B} in order to describe two slightly different subsamples of red objects of the Hubble Deep Field (HDF) North at about $z\sim0.2$ to $0.4$.
The resulting two SEDs are rather similar, such that one of them could, in principle, be dropped in a photometric redshift analysis.
In this case, however, photometric redshifts have shown to be less precise than for the case where both models have been used.
For the success of photometric redshift estimates, it is relevant that the template SEDs cover, or at least span the whole color range.
This is ensured by the Mannucci SED being too red and the other two SEDs being bluer regarding the \br\ (i.e., the $g-r$ color for $z\lesssim0.35$; see Figure~\ref{fig:ZvsCOL_std}).\\
Inspecting the SED colors in more detail (Figure~\ref{fig:ZvsCOL_std}) shows that, indeed, none of them describes the data equally well at all redshifts.
Those created for the HDF are too blue in their $g-r$ colors for $z\gtrsim0.1$, and also in $r-i$ and $i-z$ throughout the whole $z$ range.
The Mannucci S0 SED on the other hand is too red in $g-r$ at $z\lesssim0.35$, and in $r-i$ at $z\gtrsim0.15$.
It describes the observed $i-z$ color sufficiently well at $z\lesssim0.25$.
All three models match the data in $u-g$ within the errors, except for $z\sim0.15$.\\
The reason why these three SEDs give roughly correct redshifts for red galaxies is that the $g-r$ color has the strongest redshift sensitivity.
Both ellipticals predict approximately correct values of $g-r$ for $z\lesssim0.15$, whereas the Mannucci S0 matches the data well for $z\gtrsim0.32$.
For $z$ that lie in between those ranges, red galaxies are either preferentially fitted by an elliptical or the Mannucci S0 SED, depending on the photometric errors and weights applied to the $i$ and $z$ data.
For ground based data that are not as well calibrated as SDSS, the $i$ and $z$ band photometric errors are often scaled up to allow for an imprecisely known amount of telluric absorption.
Furthermore, if an SED template color has an offset from the true color which is almost constant over the whole redshift range (as it is the case for $r-i$ of the two elliptical SEDs), then there is no contribution to a systematically wrong redshift estimate.

\subsubsection{Models of Bruzual \& Charlot 2003}
\label{sec:BC03colors}

\begin{figure}
\includegraphics[width=0.5\textwidth]{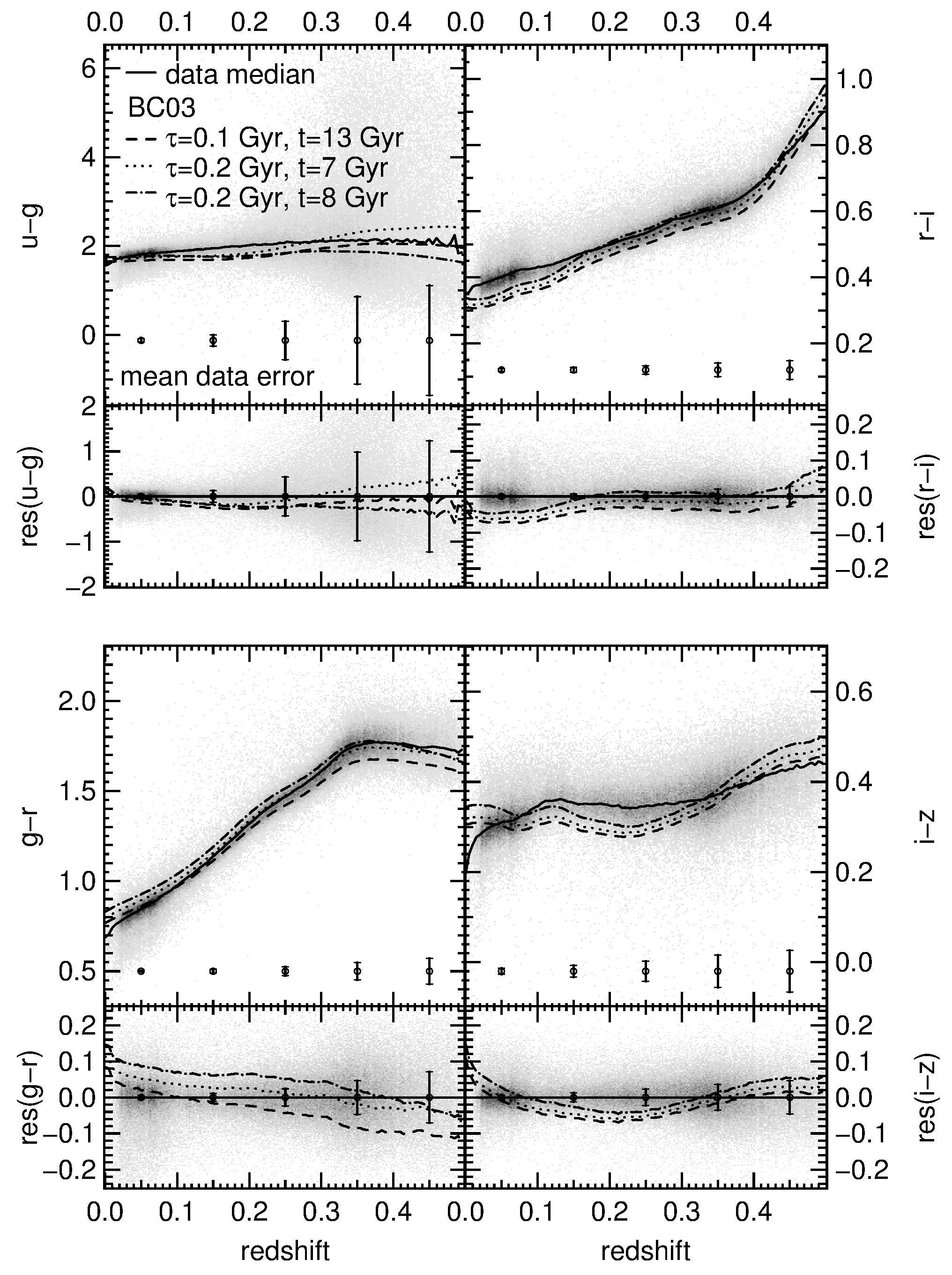}
\caption{
Upper panels: redshift vs. colors for the three BC03 models that match the colors of the LRG data best.
	  Lower panels: color residuals between models and data.
	  LRGs are displayed in gray and their median is shown by a solid line.
	  The error bars show mean data errors in five redshift bins.
	  All three models have solar metallicity, whereas  $\tau$ refers to the $e$-folding timescales of the (exponential) SFH.
	  }
\label{fig:ZvsCOL_bc03}
\end{figure}

The CSP models of \citet{2003MNRAS.344.1000B} can be generated with their software
\texttt{galaxev}\footnote{\url{http://www.cida.ve/\textasciitilde bruzual/bc2003} or \url{http://www.iap.fr/\textasciitilde charlot/bc2003}}.
We investigate SEDs from the (theoretical) BaSeL library with Padova 1994 evolutionary tracks and an initial mass function (IMF) of \citet{2003PASP..115..763C} and solar metallicity.
The models have ages from $0.01$ to $13\,\mathrm{Gyr}$, and exponential star formation rates (SFRs) with $e$-folding timescales $\tau$ between $0.5$ and $50\,\mathrm{Gyr}$ (latter introduced to account for a constant SFR).
Furthermore, we included a model that has no ongoing star formation and all stars were formed in a single burst (henceforth referred to as $\tau=0\,\mathrm{Gyr}$).
SEDs become redder in color with increasing metallicities, ages and decreasing $e$-folding timescales.
We compare only those with the data that match their median best. 
Hence, we compute the residual between the data and the solar metallicity BC03 models for the $u-g$, $g-r$, $r-i$, and $i-z$ colors.
We weighted the residuals by the data errors at redshifts greater than $0.1$ to avoid contamination by objects that do not have mostly old stellar populations:
\begin{equation}
 \mathrm{Res}^2(T)=\sum\limits_{\mathrm{objects}}\ \sum\limits_{j\in\{ ug,gr,ri,iz\} }\frac{(c_{j,\mathrm{data}}-c_{j,\mathrm{mod}})^2}{\sigma_{j,\mathrm{data}}^2}.
\label{eq:residual}
\end{equation}
The three best fitting models have $\tau=1.0\,\mathrm{Gyr}$ and ages of $t=13\,\mathrm{Gyr}$, $\tau=2.0\,\mathrm{Gyr}$ with $t=7\,\mathrm{Gyr}$ and $t=8\,\mathrm{Gyr}$.
Their colors are shown as a function of redshift in Figure~\ref{fig:ZvsCOL_bc03}.
The best fitting ages and $e$-folding timescales seem realistic, and reflect that the red colors of LRGs originate from both shorter $\tau$ and higher ages.
All three investigated models show a reasonably good match in the $u-g$ color, but are too red in $g-r$ at lower redshifts, and too blue at higher $z$.
Their $r-i$ color is bluer than the data at very small $z$ and deviates toward redder $r-i$ in the higher $z$ range. Finally, in $i-z$ they are too blue within $0.1\lesssim z\lesssim0.35$, and too red for small $z$.

\subsubsection{\bpz\ Models}

\begin{figure}
\includegraphics[width=0.5\textwidth]{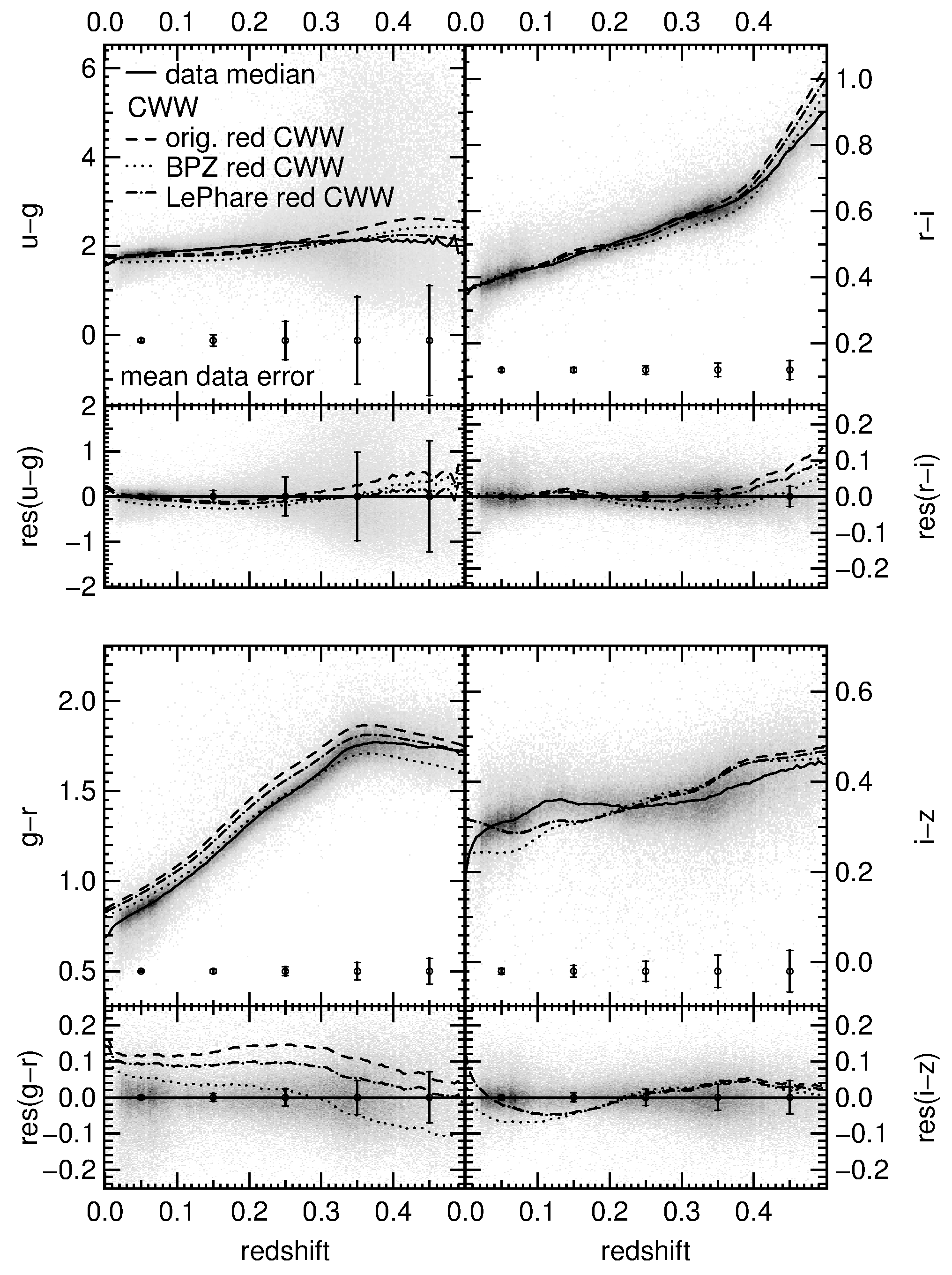}
\caption{
Upper panels: redshift vs. colors for the red CWW models: the original SED (dashed), the repaired SED of \bpz\ (dotted), and the \lephare\ CWW model (dash-dotted line).
	  Lower panels: redshift vs. color residuals between models and data.
	  LRGs are displayed in gray and their median is shown by a solid line.
	  The error bars show the mean data errors in five redshift bins.
	  }
\label{fig:ZvsCOL_CWW}
\end{figure}

The \emph{``Bayesian Photometric Redshift''} (\bpz) code\footnote{v1.98b: \url{http://acs.pha.jhu.edu/\textasciitilde txitxo/}}\footnote{v1.99.3: \url{http://www.its.caltech.edu/\textasciitilde coe/BPZ/}}
\citep{2000ApJ...536..571B, 2004ApJS..150....1B} provides a variety of template SEDs.
One is the mean spectrum of local elliptical galaxies by \citet*[][hereafter CWW]{1980ApJS...43..393C}, which was extended to the UV by \cite{2000ApJ...536..571B} through linear extrapolation, and to the near-IR with synthetic templates of \texttt{GISSEL}.
Extinction was accounted for following \cite{1995ApJ...441...18M}.
The current version of \bpz\ \citep{2004ApJS..150....1B} includes a repaired version of this elliptical SED, which was calibrated with an array of ground based data.
The colors of both the original and the repaired red CWW SED, are shown in Figure~\ref{fig:ZvsCOL_CWW}.
Both match the data well in $u-g$ and $r-i$, with the exception of the latter for higher $z$.
There are deviations from the data in $g-r$ over the whole considered $z$ range, except for $z\gtrsim0.45$ for the original CWW SED, and for medium redshifts for the repaired version.
Furthermore, they are able to match the data median in $i-z$ only for medium and high redshifts.

\subsubsection{\lephare\ Models}

\begin{figure}
\includegraphics[width=0.5\textwidth]{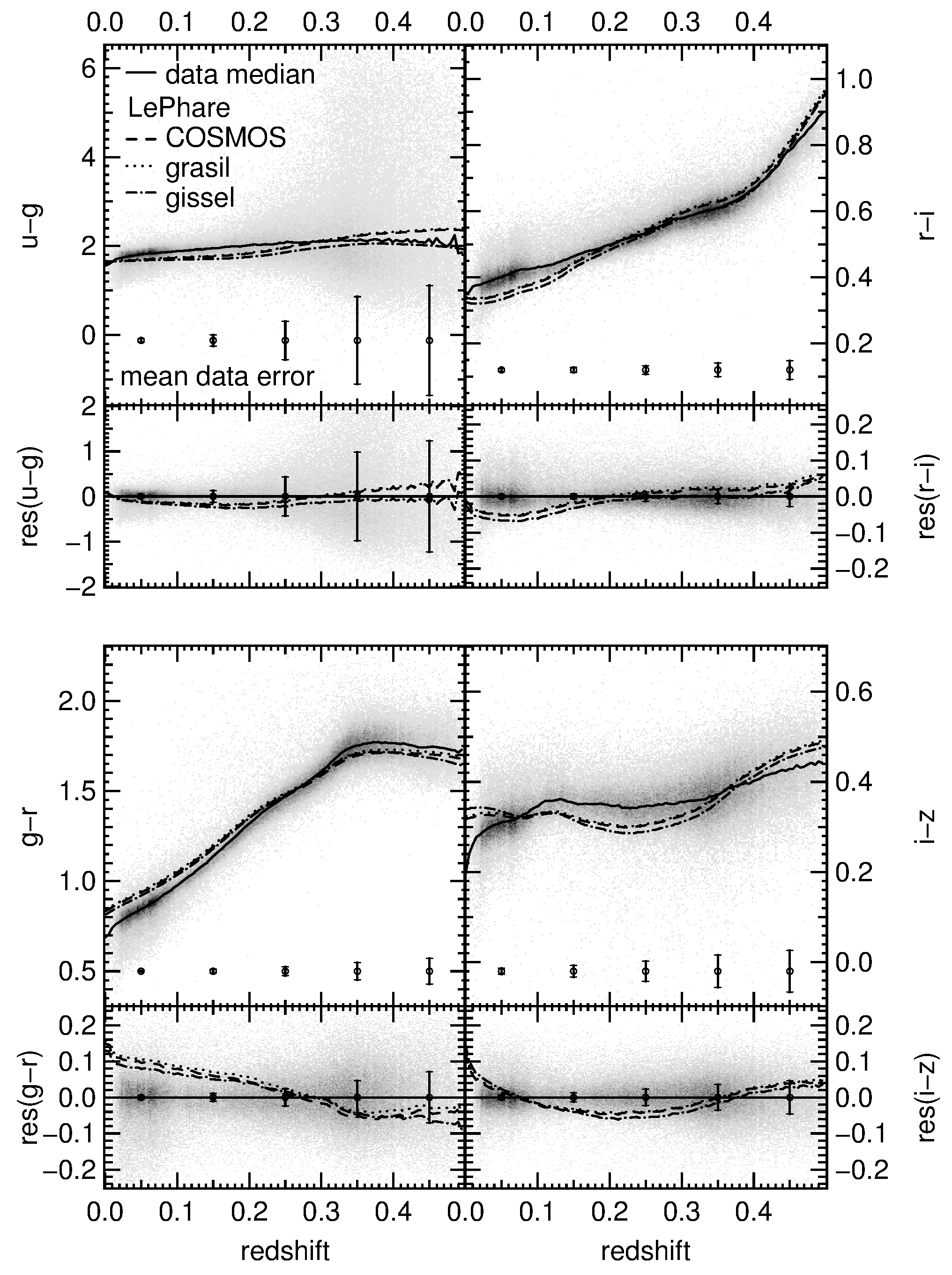}
\caption{
Upper panels: redshift vs. colors of the three models of \lephare\ that match the data best: the \emph{COSMOS} (dashed), the \texttt{GRASIL} (dotted), and the \texttt{GISSEL} SED (dash-dotted line).
Lower panels: redshift vs. color residuals between models and data.
	  LRGs are displayed in gray and their median is shown by a solid line.
	  The error bars show the mean data errors in five redshift bins.
	  A description of the model SEDs is given in the text and references therein.
	  }
\label{fig:ZvsCOL_lephare}
\end{figure}

Another Bayesian photo-$z$ software is \lephare
\footnote{\url{http://www.cfht.hawaii.edu/\textasciitilde arnouts/lephare.html}} \citep{1999MNRAS.310..540A, 2006A&A...457..841I}, which uses a wide range of template SEDs.
We adopt the nomenclature of the models used within the \lephare\ code in the following.
Again, we investigate only red SEDs that produce the smallest residuals from the data according to Eq.~\ref{eq:residual}.
The SED delivering the lowest residual originates from CWW models.
It was generated by a linear interpolation of the four CWW SEDs and a star-forming galaxy that was constructed with the population synthesis code of Bruzual \& Charlot.
The resulting colors of the interpolated red CWW SED are shown in Figure~\ref{fig:ZvsCOL_CWW} together with the colors of the original red elliptical CWW SED and its repaired version of \bpz.
Its colors are very similar to the original model and the repaired \bpz\ version, which were investigated in the last section (Figure~\ref{fig:ZvsCOL_CWW}).
The fact that these observationally created SEDs can reproduce the LRG colors well, and are therefore in principle well suited for photo-$z$ estimations, was also pointed out by \cite{2000ApJ...536..571B}.\\
Figure~\ref{fig:ZvsCOL_lephare} shows the colors of three other SEDs that yield the second, third and fourth smallest residuals from the data.
The one termed \emph{COSMOS} is from the \lephare\ set called \emph{COSMOS\_SED}.
Models from this set were used in \citet{2009ApJ...690.1236I} for photometric redshift estimates in the $2\ \mathrm{deg}^2$ COSMOS field.
The model tagged \emph{grasil} is a $13\,\mathrm{Gyr}$ old elliptical galaxy that was created with the \texttt{GRASIL} code \citep{1998ApJ...509..103S}.
The SED describing the data fourth best comes from the \lephare\ \emph{42\_GISSEL} set.
It has an $e$-folding timescale of $1\,\mathrm{Gyr}$, a metallicity of $Z=0.02\ (Z_{\sun})$ and an age of $7\,\mathrm{Gyr}$.
From Figure~\ref{fig:ZvsCOL_lephare} one can infer that the $g-r$ colors of those three models are redder than the data for low $z$.
In $r-i$, the three model SEDs are too blue for lower redshifts.
Their $i-z$ is redder than the data at $z\lesssim0.07$ and $z\gtrsim0.45$, and bluer at $0.07\lesssim z\lesssim0.3$

\subsubsection{Models of Maraston et al. (2009)}
\begin{figure}
\includegraphics[width=0.5\textwidth]{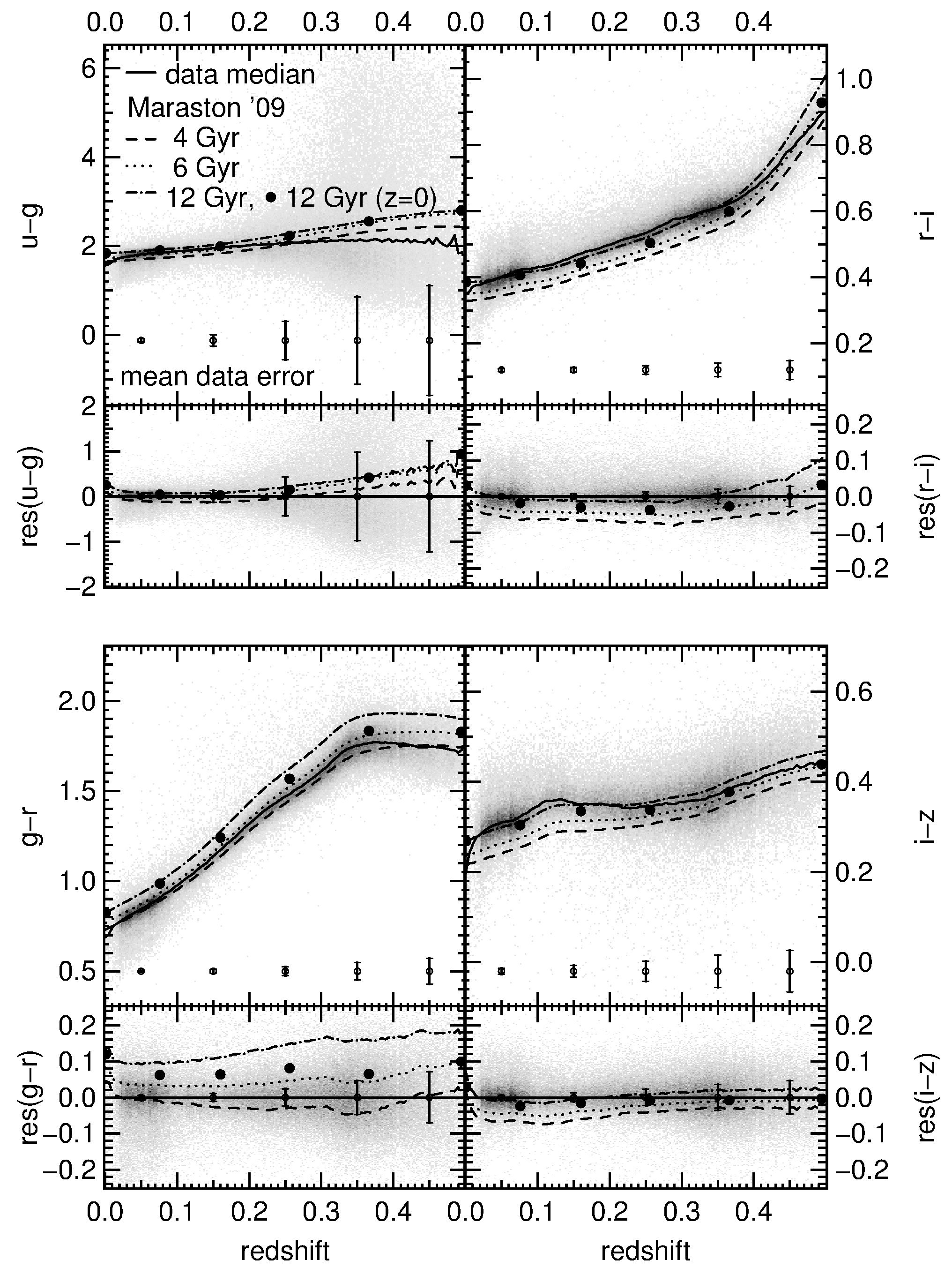}
\caption{
Upper panels: redshift vs. colors for three M09 models, where the LRG with an age of $6\,\mathrm{Gyr}$ (dotted line) matches the data best regarding the residual to the median.
	  Lower panels: redshift vs. color residuals between models and data.
	  LRGs are displayed in gray and their median is shown by a solid line.
	  Error bars show the mean data errors in five redshift bins.
	  The other lines display the M09 models that have an age of $4\,\mathrm{Gyr}$ (dashed), and $12\,\mathrm{Gyr}$ (dash-dotted line).
	  The solid circles show a $12\,\mathrm{Gyr}$ (at $z=0$) old M09 SED model evolving with time.
	  The following dots represent the same star formation model, but with ages of $11$, $10$, $9$, $8$, and $7\,\mathrm{Gyr}$.
	  A description of the model SEDs is given in the text and references therein.
	  }
\label{fig:ZvsCOL_M09}
\end{figure}

\citet{2009MNRAS.394L.107M} already addressed the problem of the color mismatch between model SEDs for LRGs and the actual data.
They studied several SEDs and compared them to the data median, focusing on $g-r$ and $r-i$, and discovered that neither the synthetic nor the empirical SEDs could accurately represent the data at all redshifts equally well.
To produce new models that resemble the color median of the data more precisely, Maraston et al. added a mass fraction of 3\% blue metal-poor stars to create empirically corrected CSP models.
It enabled them to match the data colors considered in their publication ($g-r$ and $r-i$) very well.
We present the colors of a subset of the M09 model SEDs\footnote{\url{http://www.icg.port.ac.uk/\textasciitilde maraston/Claudia's\_Stellar\_Population\_Model.html}} in the $u-g$, $g-r$, $r-i$, and $i-z$ colors in Figure~\ref{fig:ZvsCOL_M09}.
The models that are displayed by lines have not been evolved with look-back time.
The SED with an age of $6\,\mathrm{Gyr}$ yields the smallest residual (according to Equation~\ref{eq:residual}) to the data.
It increases faster when turning to younger populations than it does for older ages.
In order to show the behavior for ages different from $6\,\mathrm{Gyr}$, we display one older ($12\,\mathrm{Gyr}$) SED in Figure~\ref{fig:ZvsCOL_M09}, and one younger model with an age of $4\,\mathrm{Gyr}$.\\
In contrast to the SEDs examined above, the M09 redshift\textendash color relation has the same shape as that of the data median.
The $u-g$ color deviation of the M09 models from the data lies well below the mean error, and does not differ notably for the models with $6\,\mathrm{Gyr}$ and $12\,\mathrm{Gyr}$.
At $z\lesssim0.2$ the $u-g$ color of the $4\,\mathrm{Gyr}$ SED resembles that of the other two SEDs but has lower values for higher redshifts.
The $g-r$ color of the $6\,\mathrm{Gyr}$ old LRG is close to the data at low $z$ ($\lesssim0.25$) but deviates at $z\gtrsim0.35$.
Its $r-i$ and $i-z$ colors are too blue in comparison with the data at $z\lesssim0.4$ and $z\lesssim0.3$ respectively.
The $12\,\mathrm{Gyr}$ old SED matches the data in $r-i$ well for $z\lesssim0.4$ and in $i-z$ for all $z$, whereas its $g-r$ color is too red in the whole redshift range.
The LRG with an age of $4\,\mathrm{Gyr}$ is bluer with respect to the data in $r-i$ at all redshifts and $i-z$ at $z\lesssim0.3$, as well as in $g-r$ at intermediate redshifts ($0.2\lesssim z\lesssim0.35$).
Unsurprisingly, the older SEDs represent the data at lower redshifts better, while the younger population exhibits colors matching better at higher $z$.
An exception is $g-r$ where the oldest model shows an offset from the data, whereas the $6\,\mathrm{Gyr}$ old model fits well up to $z\approx0.35$.
The SED with an age of $4\,\mathrm{Gyr}$ matches the data reasonably well in $g-r$ throughout the whole redshift range.\\
Figure~\ref{fig:ZvsCOL_M09} shows furthermore the M09 model with an age of $12\,\mathrm{Gyr}$ at $z=0$ evolving with look-back time.
The dots at higher $z$ represent ages of $11$, $10$, $9$, $8$, and $7\,\mathrm{Gyr}$.
The colors of the redshift-evolved M09 model are in very good agreement with the data compared to the previously investigated model SEDs.
However, the model which is $12\,\mathrm{Gyr}$ old at $z=0$ and evolves with redshift exhibits an overall offset from the data median in $g-r$, and has a bluer $r-i$ color for $0.15\lesssim z\lesssim0.35$.
Furthermore, it deviates in $u-g$ at higher redshifts ($z\gtrsim0.3$), but still lies well within the errors.

\subsubsection{Summary}
Altogether, we conclude that we could not find models that are able to represent the SDSS LRG data within their errors at all redshifts.
The reason for this is that the SEDs differ from the data median as a function of $z$.
The M09 models (not taking evolution with look-back time into account; shown with lines in Figure~\ref{fig:ZvsCOL_M09}) describe the SDSS LRGs best compared to the templates examined before, but they also suffer from notable deviations within some redshift intervals.
The lower redshift range ($z\lesssim0.1$) and the $i-z$ colors were not considered in \citet{2009MNRAS.394L.107M}.\\
As expected, the colors of the redshift evolved SED with an age of $12\,\mathrm{Gyr}$ at $z=0$ (solid circles in Figure~\ref{fig:ZvsCOL_M09}) match the data better as a function of redshift, although not perfectly.
Furthermore, if only the age of a SED is varied, one just obtains a one-dimensional sequence in color\textendash color space.
Instead, the colors of true SEDs are spread in two dimensions, which is a clear hint that not only the age of the LRGs, but also the SFH, and maybe metallicity, vary.
A wider spread can thus only be produced if different SFHs and metallicities are considered, which are not publicly available for M09 models.
We therefore conclude that in order to get SEDs which represent the data at all redshifts and in all colors, one has to consider different SED template sets for different redshift bins.

\section{New SED Templates}
\label{sec:sedfitmodels}
In the past, optimal template SEDs have been obtained by ``repairing'' individual SEDs empirically using the mismatch of observed and model colors \citep[e.g.,][]{2003AJ....125..580C}.
In addition, or alternatively, a larger variety of template SEDs has been obtained by interpolating between a smaller number of template SEDs \citep[e.g.,][]{2008ApJ...686.1503B}.
However, there are some caveats. If models that one starts with are a bad match to the data it is unlikely that a combination of them will be any better.
Moreover, if one introduces SEDs that do not match the data, one deteriorates the photometric redshift quality.
Furthermore, the question of how many SEDs are needed to match the data remains undecided.

\subsection{Generating Best Fitting SEDs for Individual Objects in the LRG Catalog by SED Fitting}
\label{sec:generation}
We start with SEDs that match the photometry of individual galaxies as well as possible, and subsequently select a small subset of these SEDs with the goal that they together describe the whole data set.
To derive SEDs for individual galaxies, we make use of the largest appropriate freedom to describe a SED with five band data.
I.e., we describe the SED of an LRG by a superposition of a CSP and a burst model.
Furthermore, we allow for extinction of the CSP and the burst component.
The burst and its extinction are physically motivated by the fact that LRGs can contain some young stars.
The extinction of the CSP is mainly there to increase the degree of freedom and allows to change the continuum slope (i.e., to introduce a ``variance'' of CSP models) of the main stellar population.\\
We generate SEDs for LRGs with the SED fitting routine \sedfit\ \citep{2004ApJ...616L.103D}.
It chooses the best fitting CSP+burst combination through $\chi^2$ minimization, concurrently allowing for dust reddening of both components following the extinction law of \cite{2000ApJ...533..682C}.\\
In Section~\ref{sec:modelcolors} we pointed out that for each model there is always a discrepancy between the model and data colors at some $z$, and we are not able to pick a set of models which describe the data equally well at all $z$.
Thus, we create templates specifically designed to represent different redshift regions.
We do so by splitting the spectroscopic SDSS LRG sample into five subcatalogs, all of them containing objects within redshift intervals of width $0.04$, centered on $z=0.02$, $0.1$, $0.2$, $0.3$ and $0.4$.
These catalogs contain $9660$, $38{,}583$, $9292$, $15{,}093$, and $11{,}667$ objects.
They are then divided into two equally sized subcatalogs; one serving for template generation (derivation half) and the other one serving as reference when estimating photometric redshift accuracies (validation half).
Subsequently, we run \sedfit\ with a model set consisting of 864 BC03 models from the BaSeL library \citep{2003MNRAS.344.1000B} with Padova 1994 evolutionary tracks and an IMF from \citet{2003PASP..115..763C}.
They have ages between $0.01$ and $13\,\mathrm{Gyr}$, $e$-folding timescales of $\tau=0.0$, and $\tau=0.5$ to $50\,\mathrm{Gyr}$, and four different metallicities, $Z=0.004$, $0.008$, $0.02\ (Z_{\sun})$ and $0.05$.
In order to fit the photometry, these models can be superposed by a fraction of the burst model, with an essentially constant SFR (i.e., $\tau=20\,\mathrm{Gyr}$), solar metallicity, and an age of $50\,\mathrm{Myr}$.
The burst mass-fraction is constrained to $\left[0.0,0.01\right]$ with steps of $0.002$.
Both the CSP and the burst component can be extincted separately.
This fitting procedure results in a SED that reproduces the data as well as possible.
We end up with one ``best fitting SED'' for each object in the template generation catalog.\\
The relative performance of SED fitting with models of BC03, M09, and \citet[hereafter M11]{2011MNRAS.418.2785M} is demonstrated in Appendix~\ref{app:models}, where we also justify why we prefer BC03 models over M09/11 for our SED construction procedure.
From Figure~\ref{fig:COLSEDfit} in Appendix~\ref{app:models} we conclude that BC03 models allow a better coverage of the colors of the LRGs than M09/11 models do, and fit the data better (Figure~\ref{fig:SEDfitchi2}).

\subsection{Selection of Best Fitting SEDs for the New Template Set}
\label{sec:sedfitmodelselection}
\begin{figure}
\centering
\includegraphics[width=0.45\textwidth]{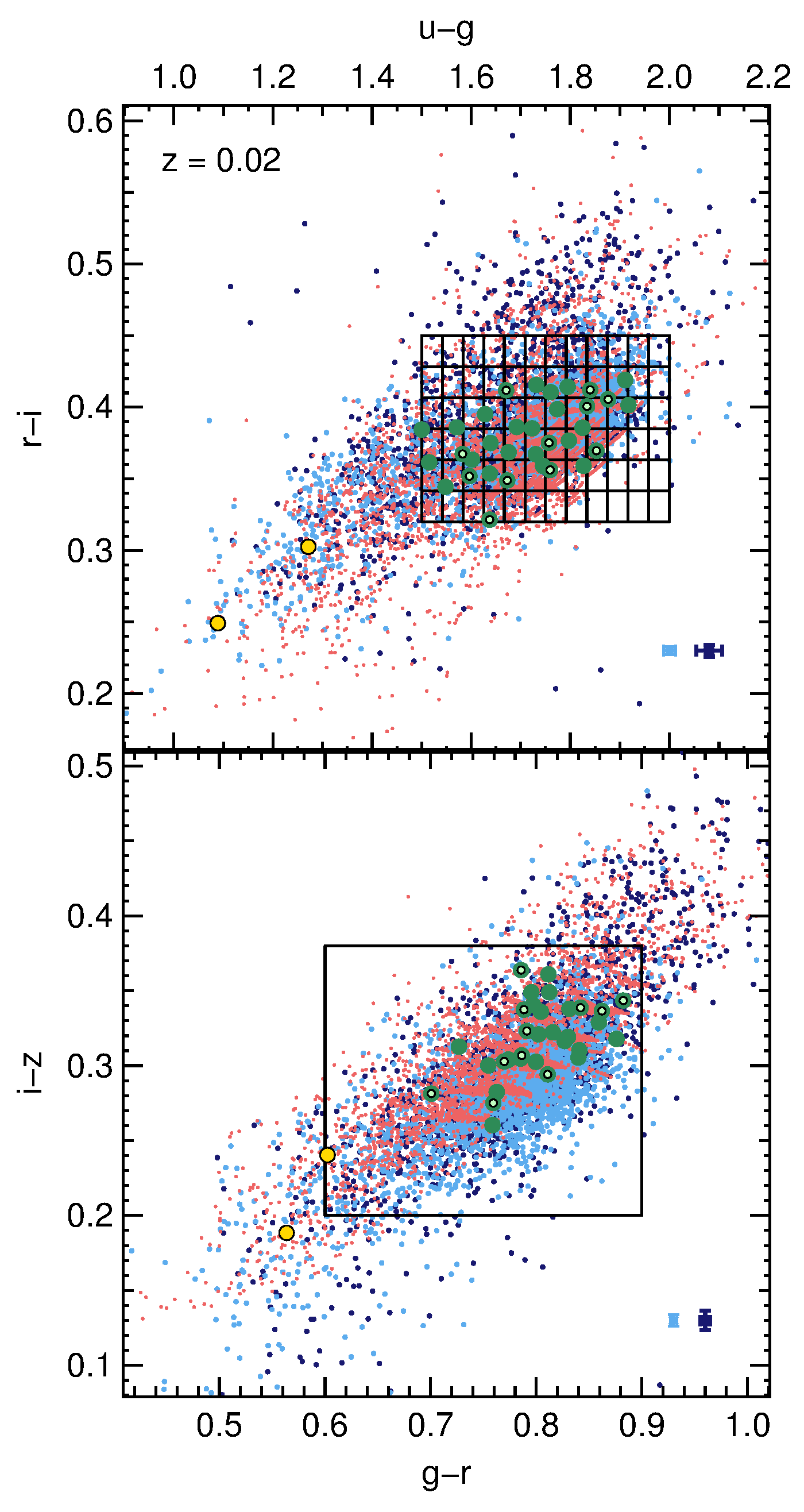}
\caption{Color vs. color plots for SDSS data (blue) and their individual best fitting \sedfit-SEDs (red) for $0.0\leq z\leq0.04$.
	  Objects with $u$ band errors lower than the median are indicated in light blue, whereas those with greater errors are dark blue.
	  The grid (upper panel) and the boundary (lower panel) within which the models are selected are shown in black.
	  The dark green dots are the preselected models, whereas the light green points represent the models that are left over after the removal of redundant SEDs.
	  The yellow dots are models that shall account for objects outside the selected boundaries in $ugri$ and $griz$.}
\label{fig:selectbycol_z00}
\end{figure}

We want to select a set of models from all the SEDs we produced for the template generation subcatalogs such that this set represents the five redshift subcatalogs in color space.
We explain our procedure for the $z\approx0$ redshift interval using Figure~\ref{fig:selectbycol_z00}.

\begin{figure}
\includegraphics[width=0.5\textwidth]{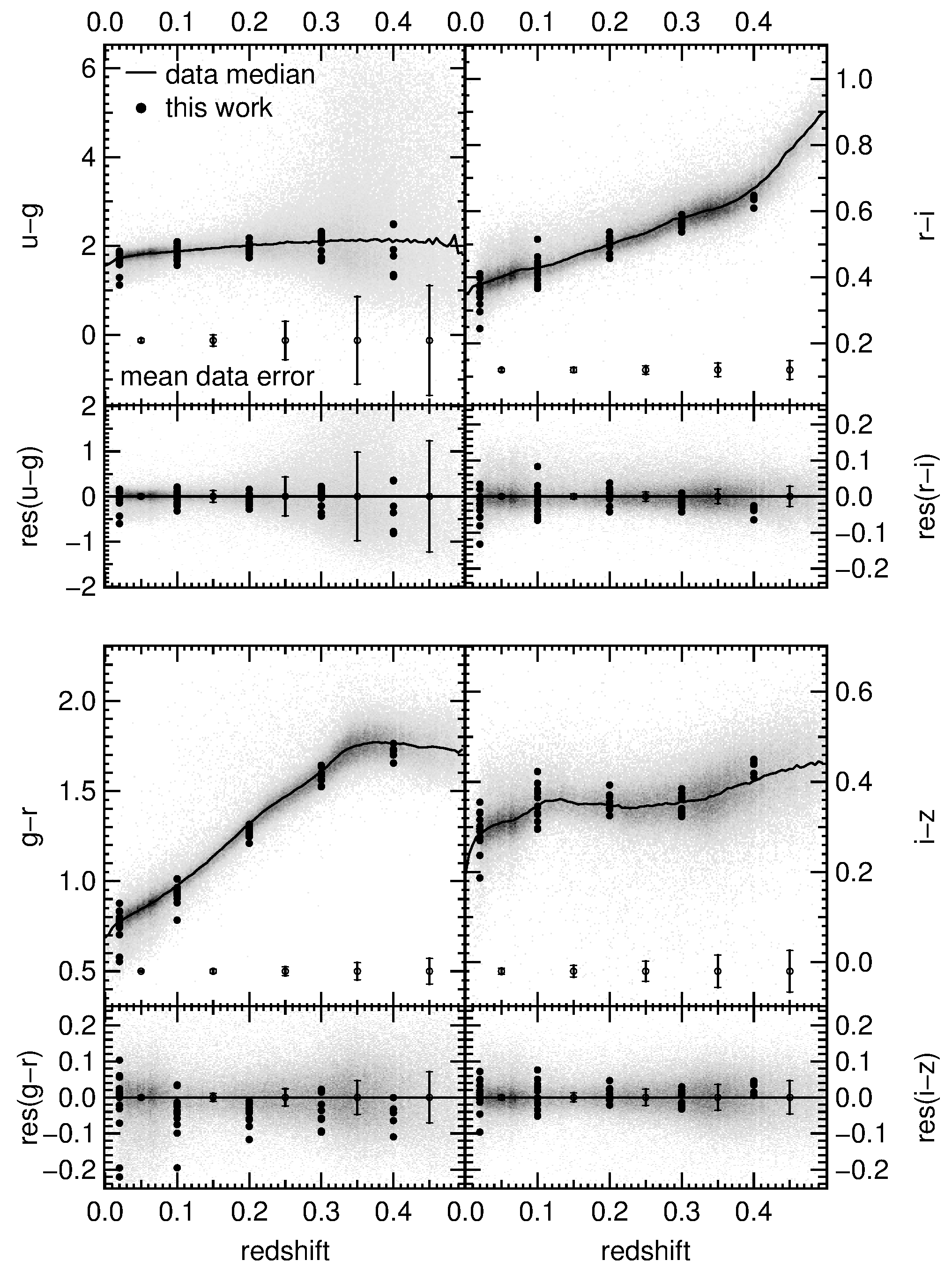}
\caption{
Upper panels: colors of the novel template SEDs as a function of redshift.
	  Lower panels: redshift vs. color residuals between models and data.
	  LRGs are represented by the density plot, whereas the solid line is their median.
	  The error bars show the mean data errors in five redshift bins.
	  The colors and the color residuals of the new template SEDs are represented by points.
	  They are plotted at the mean values of the redshift bins from which they were created.
	  }
\label{fig:ZvsCOL_SEDfit}
\end{figure}

First we plot the distributions of the measured colors of the galaxies in the $u-g$ versus $r-i$ plane.
Objects which have a $u$ band error smaller (larger) than the median are displayed with light (dark) blue points.
Second, we plot the colors of the best fitting SED for each object with red points.
If the best fitting SEDs are a good match to the galaxy data, then the distribution of them in color space should be similar to the observed distribution, once the spreading by photometric errors is taken into account.
The mean photometric errors for the two subsamples (with high and low $u$ band errors) are shown in dark and light blue in the lower right corner of the panels in Figures~\ref{fig:selectbycol_z00}, and \ref{fig:selectbycol_z00-2}\textendash\ref{fig:selectbycol_z04}.
It shows that galaxies with larger photometric errors in $u$ in general also have larger photometric errors in $r-i$ and $i-z$, which implies that these objects are fainter in all observed bands, rather than just in the $u$ band.
We conclude from the shown sizes of the mean photometric errors on one hand that the spread in colors is of physical origin, and not due to photometric errors.
This also holds for higher redshifts where the error bars are of significant size.
In order to confirm that, we simulated galaxy colors by assuming that the spread in color space is a result of the photometric errors only.
We simulate the data by assuming that each object in fact has a color value of the average color.
Then we assign a random deviation from this point, based on a Gaussian probability distribution with the measured error as standard deviation.
Since the spread in the simulated data is smaller than that of the data, we confirm that the spread within the colors is of physical origin.
On the other hand, we infer that the color distribution should hardly be broadened by photometric errors, and that the best fit SED color reproduce the variety of 'true' galaxy colors well.
Only for the reddest colors ($u-g>1.5$ and $r-i>0.45$) we find mostly objects with larger than median photometric errors, such that some of these data points will be scattered out from the true distribution by large photometric errors.\\
We now select model SEDs for the ``main galaxy'' population.
Therefore, we define an area in the $ugri$-plane which contains the majority of objects (black box), and then set a grid within this area (black grid).
The chosen area is selected by eye and was confirmed to include at least $70\%$ of the objects from that redshift bin.
We want to select one object per cell which shall represent the other objects within that cell.
We carry out this procedure only for cells that contain at least 20 galaxies.
From each of those cells we pick the SEDs best fitting the five objects from that cell which have the lowest \emph{u} band error.\\
As mentioned above, our objective is to produce photometric redshifts with biases as small as possible, and we therefore try to single out SEDs that produce small $\langle\dz\rangle$ on the catalogs of the regarded $z$ range.
Hence, we perform \photoz\ runs on the derivation half and on the validation half of the $z\approx0.02$ subcatalog, the latter to confirm that we did not unintentionally divide the catalog into halves with different object properties.
We perform the photometric redshift estimation with each of the preselected models separately, thus fitting only one model per run.
Afterward, we pick the SED that renders the smallest bias on both subcatalog halves out of the five originating from the same cell.
The selected model shall represent the objects within its neighborhood in color space in the following \photoz\ runs.
Repeating the same procedure for all cells and redshift bins, we select roughly $20$ to $35$ model SEDs per $z$ bin.\\
In order to save computation time for a \photoz\ run, we want to keep the number of templates to a minimum.
Therefore, we calculate the root mean squared flux difference of each SED combination at every defined wavelength.
The most similar SEDs are then removed from the final set, leaving typically about $10$ models per redshift bin with a total number of $49$ templates.\\
The redshift of the models is plotted versus their color in Figure~\ref{fig:ZvsCOL_SEDfit}.
Color\textendash color relations of the new models in comparison to the LRGs are displayed in Figure~\ref{fig:selectbycol_z00}, and Appendix~\ref{app:colcol}, Figures~\ref{fig:selectbycol_z00-2}\textendash\ref{fig:selectbycol_z04}.
The templates represent the LRG colors and their spread for most values of $z$.
The exceptions at $z\sim0.4$ are discussed in Appendix~\ref{app:models}, Figure~\ref{fig:selectbycol_z04}.
We test the performance of the new models with respect to photometric redshifts in the next section.\\
As an alternative we also selected template SEDs with a kd-tree.
We generated one hyperplane for each of the four colors $u-g$, $g-r$, $r-i$, and $i-z$, thus a four-dimensional tree with 16 template SEDs for each redshift bin.
One model was selected from each branch of the tree by the same criteria as already described.
As this technique yielded SEDs not too much different, we kept our hand-selected set.

\section{Photometric Redshift Precision with the Novel Template SEDs}
\label{sec:results}
\begin{figure}
\includegraphics[width=0.45\textwidth]{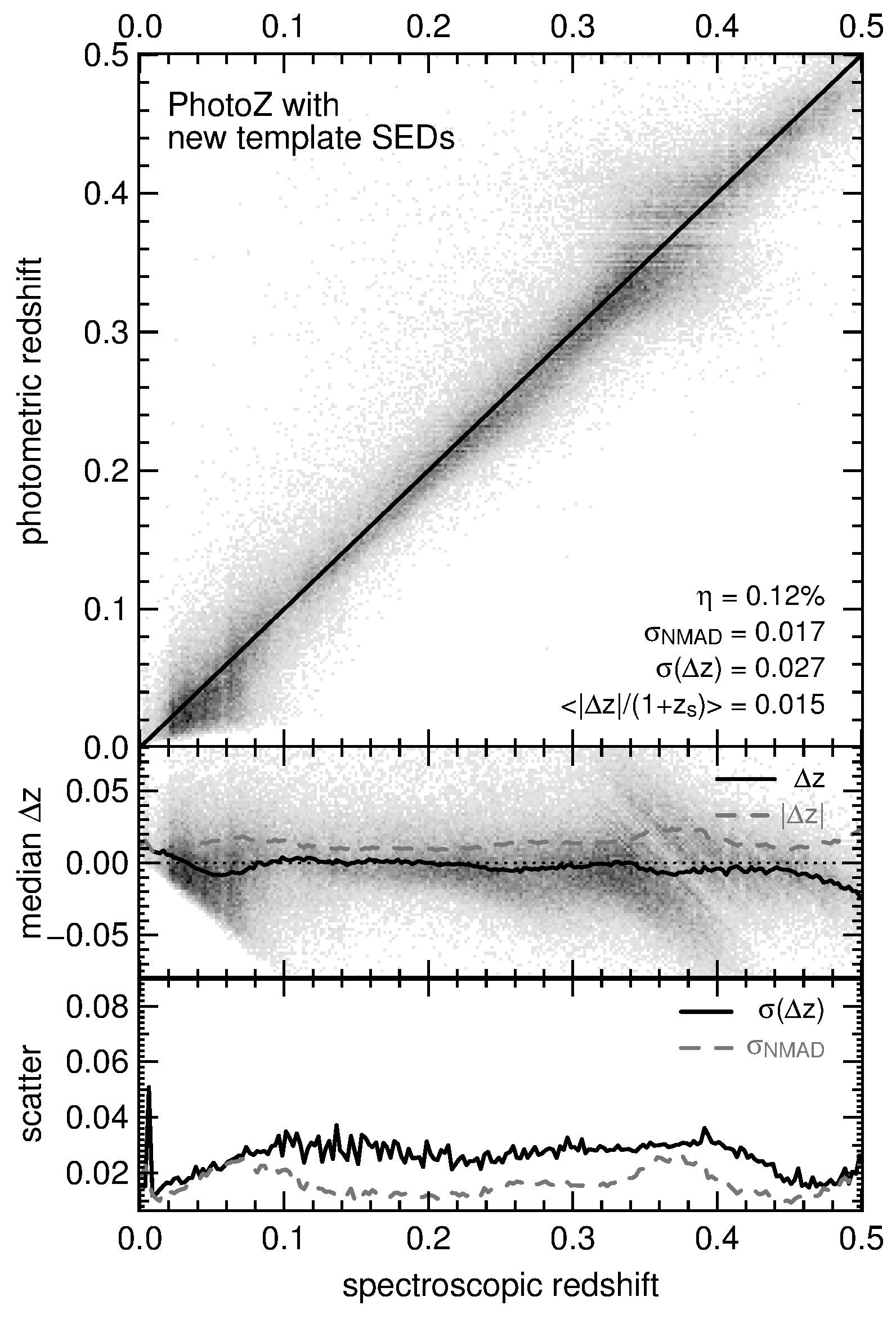}
\caption{Photometric redshift results of a \photoz\ run with the newly created templates and $140{,}331$ SDSS LRGs.
We removed $20$ objects that had undefined photometric redshifts.
$\sigma_{\Delta z}$, $\snmad$, and $\eta$ are explained in Section~\ref{sec:results}.}
\label{fig:ZSvsZP_SEDfitred}
\end{figure}

Before analyzing the photometric redshift results we get with the new template SEDs, we introduce some quality parameters.
To define badly estimated photometric redshifts we use the threshold $|\zphot-\zspec|(1+\zspec)>0.15$ introduced by \citet{2006A&A...457..841I}.
Objects that fulfill this inequality are tagged ``catastrophic outliers'', and their number fraction will be called $\eta$.
$\sigma_{\dz}$ shall denote the root mean square of the photometric redshift error $\dz=\zphot-\zspec$.
A parameter that quantifies the distribution perpendicular to the $\zspec=\zphot$ line without
the inclusion of catastrophic outliers is $\snmad=1.48\times \mathrm{median}\left(\frac{|\dz|}{1+\zspec}\right)_{\mathrm{non-outliers}}$ \citep{2006A&A...457..841I}.
With $\snmad$ one can make conclusions about the redshift uncertainty excluding the tails of the distribution.\\
After these definitions we turn to the integration of the newly created model SEDs into \photoz.
In Section~\ref{sec:PhotoZ} we mentioned the redshift priors we can impose on each template.
The total probability for a model-$z$ combination is proportional to the redshift probability $P_z$.
As already mentioned in Section~\ref{sec:PhotoZ}, we use a Gaussian probability distribution, $P_z(z)\approx\exp\left(-\ln2\left(\frac{z-\hat{z}}{0.2}\right)^2\right)$.
In this way, $\hat{z}$ can be set to the according redshift interval's center for each model, thus $\hat{z}=0.02,0.1,0.2,0.3,0.4$.
The redshift prior probability of the model is equal to \onehalf\ at the corresponding $\hat{z}\pm0.2$.\\
The photometric redshift results are shown in the upper panel of Figure~\ref{fig:ZSvsZP_SEDfitred}.
With the new red templates, the described priors, and a resolution of $0.002$ we indeed get very accurate redshifts.
The catastrophic outlier rate is very small at $\eta=0.12\,\%$.
The dispersions is $\sigma_{\dz}=0.027$ and $\snmad=0.017$ respectively.
The mean absolute deviation of $\zphot$ from $\zspec$ is $\langle|\dz|/(1+\zspec)\rangle=0.015$.
We show the bias as a function of redshift in the middle panel of Figure~\ref{fig:ZSvsZP_SEDfitred}.
It exceeds $0.01$ only for the highest redshifts ($z\geq0.47$).
The reason for the small positive bias at $z\lesssim0.02$ is that there are no negative photometric redshifts that can cancel out overestimation.
The total mean bias reads $-0.0004$.
At $\zspec\gtrsim0.36$, the \br\ position in wavelength cannot be determined to high precision.
This is because at $z\sim0.36$ the \br\ is observed at $\lambda\sim5500\,\textrm{\AA{}}$ and thus lies in the gap between the SDSS $g$ and $r$ filter.
We therefore expect that the error distribution widens at this redshift.
The reason for the error becoming bimodal is likely because at this redshift we do not have ``perfect templates'' and thus the photometric redshifts are either over- or underestimated.
Luckily the errors nearly average out for every redshift bin.\\
The bottom panel of Figure~\ref{fig:ZSvsZP_SEDfitred} shows the dependence of the scatter $\sigma_{\dz}$ and $\snmad$ on $z$.
At low $z$ ($\lesssim0.01$) it is up to $0.04$, which is likely due to contamination of the LRG sample with non LRGs.
This is supported by the fact that the $\chi^2$ values in this region are significantly higher than for $z\gtrsim0.01$.
Apart from very small redshifts, the scatter stays more or less the same for all redshifts.

\section{Comparison to SDSS Database Results}
\label{sec:comparison}
\begin{figure}
\includegraphics[width=0.45\textwidth]{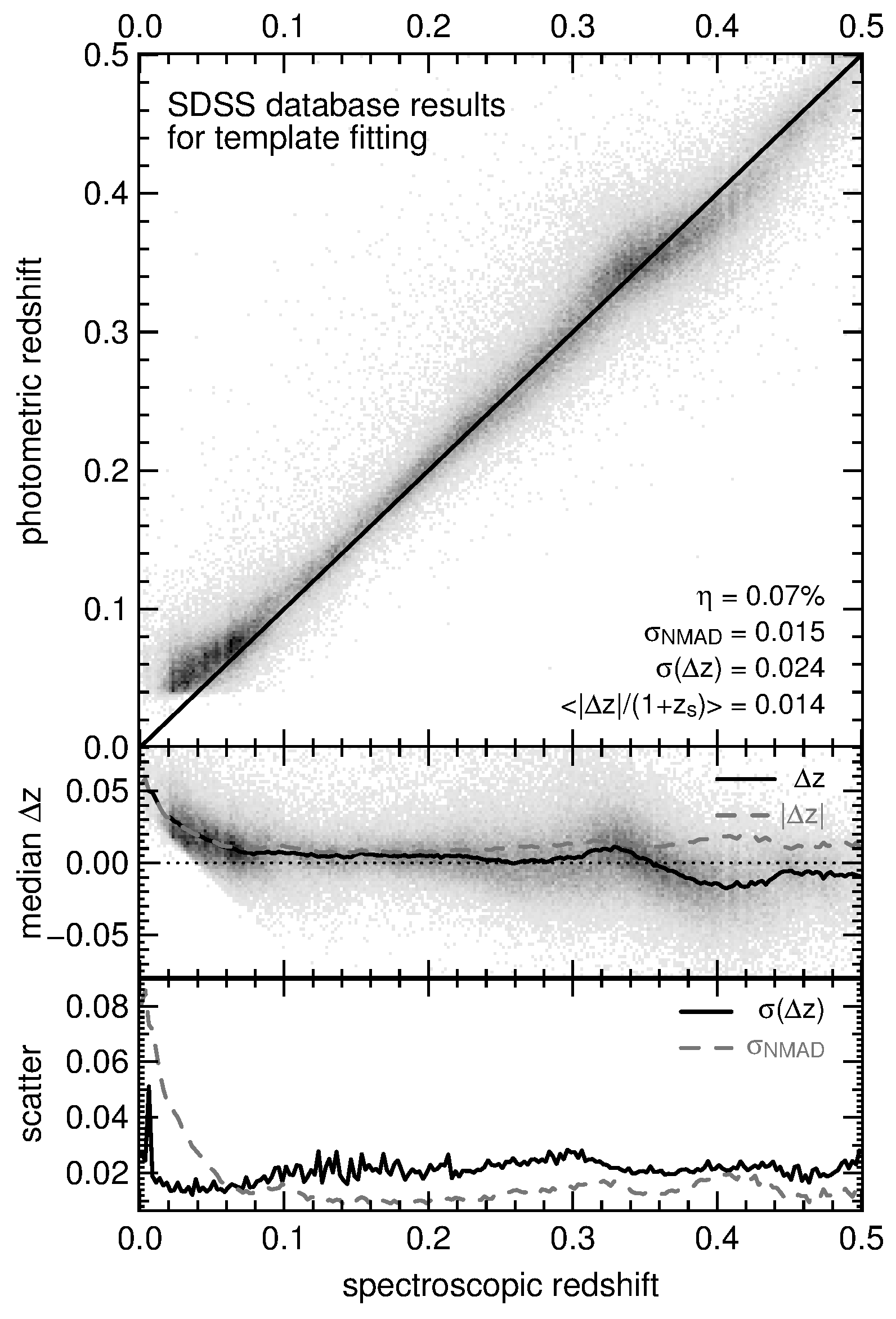}
\caption{Photometric redshift results of SDSS template fitting.
We removed those objects with undefined photometric redshifts.
$\eta$, $\snmad$, and  $\sigma_{\dz}$ are explained in Section~\ref{sec:results}.}
\label{fig:ZSvsZP_SDSSfitting}
\end{figure}

\begin{figure}
\includegraphics[width=0.45\textwidth]{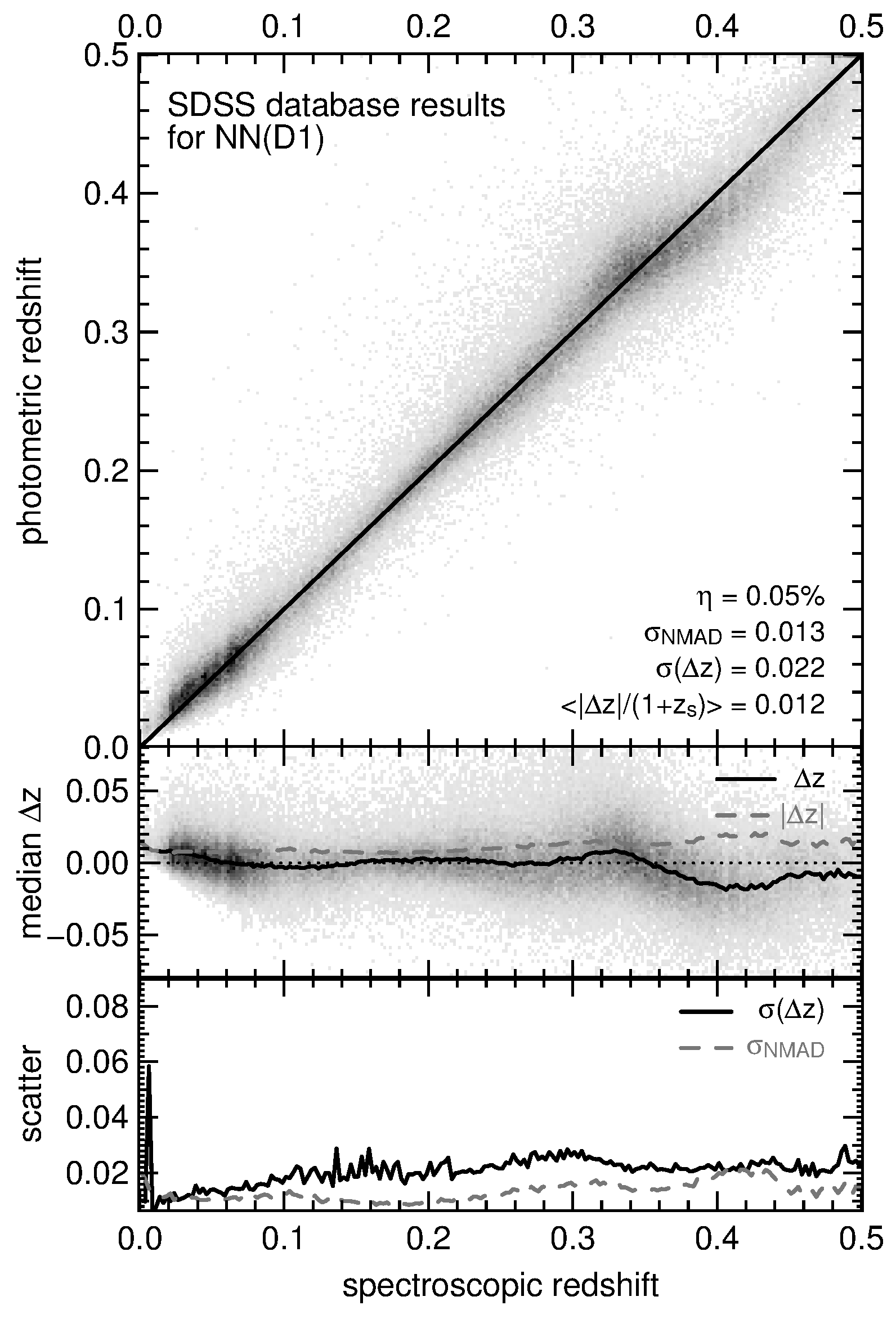}
\caption{Photometric redshift results of SDSS ANN(D1).
We removed those objects with undefined photometric redshifts.
$\eta$, $\snmad$, and  $\sigma_{\dz}$ are explained in Section~\ref{sec:results}.}
\label{fig:ZSvsZP_SDSSd1}
\end{figure}

\begin{figure}
\includegraphics[width=0.45\textwidth]{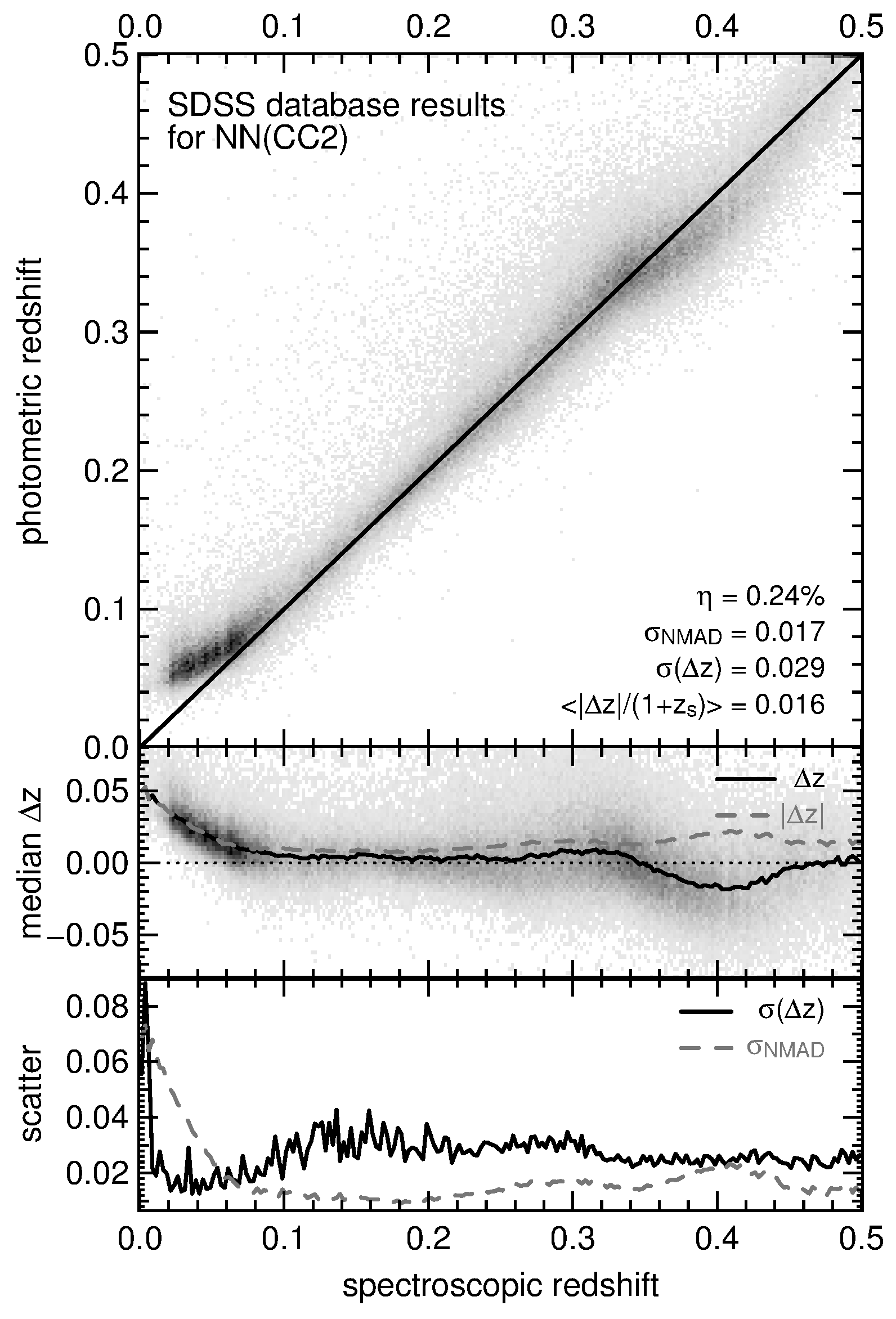}
\caption{Photometric redshift results of SDSS ANN(CC2).
$\eta$, $\snmad$, and  $\sigma_{\dz}$ are explained in Section~\ref{sec:results}.}
\label{fig:ZSvsZP_SDSScc2}
\end{figure}

We now compare our photometric redshift results with those in the SDSS database.
There are three kinds of photometric redshifts available in the database: One was obtained through template fitting (i.e., a hybrid technique of template fitting and a reparation of templates, see \citealt{2003AJ....125..580C}), and the other two are results of an artificial neural network (ANN) approach \citep{2008ApJ...674..768O}.
The results of the template fitting are shown in Figure~\ref{fig:ZSvsZP_SDSSfitting}.
On average, the $\sigma_{\dz}$ and $\snmad$ of these redshifts are smaller than for our method, but the bias is more unevenly spread in redshift space.
Also for this method the median bias is smaller than $0.01$ only in the redshift range of $0.07\leq z\leq0.32$.
For other redshifts the bias exceeds that of our redshifts considerably.\\
The SDSS database values for photometric redshifts obtained with the ANN are shown in Figures~\ref{fig:ZSvsZP_SDSSd1} and \ref{fig:ZSvsZP_SDSScc2}, where we display the results of two techniques, \emph{D1} and \emph{CC2}.
The \emph{ANN(D1)} method uses \emph{ugriz} magnitudes and the concentration indices of all five bands for the determination of photometric redshifts.
A concentration index of a passband $i$ is defined as the value of the ratio between the radii encircling $50\%$ and $90\%$ of the Petrosian flux.
The \emph{CC2} approach on the other hand is based on the $u-g$, $g-r$, $r-i$, and $i-z$ colors and the concentration indices in the \emph{g}, \emph{r} and \emph{i} bands.
Photometric redshift results are shown in Figures~\ref{fig:ZSvsZP_SDSSd1} (\emph{D1}) and \ref{fig:ZSvsZP_SDSScc2} (\emph{CC2}), respectively.\\
The \emph{ANN(D1)} performs excellently in terms of scatter and bias from $z=0.01$ to $z=0.3$, but it yields a considerably larger bias than our redshifts for $z>0.3$.
Photometric redshifts from the \emph{CC2} method are significantly overestimated for $z\leq0.08$, where the bias exceeds $0.01$.
Additionally, the $\snmad$ is particularly high at small spectroscopic redshifts ($z\leq0.06$), which is probably due to the overestimation in this range.
At $z\approx0.3$, $\dz$ surpasses $0.01$, whereas at intermediate redshifts, the photometric redshifts from the \emph{ANN(CC2)} are slightly overestimated.\\
We conclude that \photoz\ with the new set of LRG templates delivers results competitive with empirical approaches.

\section{Properties of the Novel Template Set}
\label{sec:discussion}
In this section we want to discuss the properties of our new template SEDs in the UV wavelength range and analyze how well they represent objects with UV data.
Being able to match the UV colors of observed objects justifies that we discuss the novel SEDs from a physical point of view and with regard to the underlying galaxy population in Section~\ref{sec:zvsmabs}.

\subsection{UV colors of the Novel Templates and M09 Models}
\label{sec:UV}

\begin{figure*}
\centering
\includegraphics[width=0.8\textwidth]{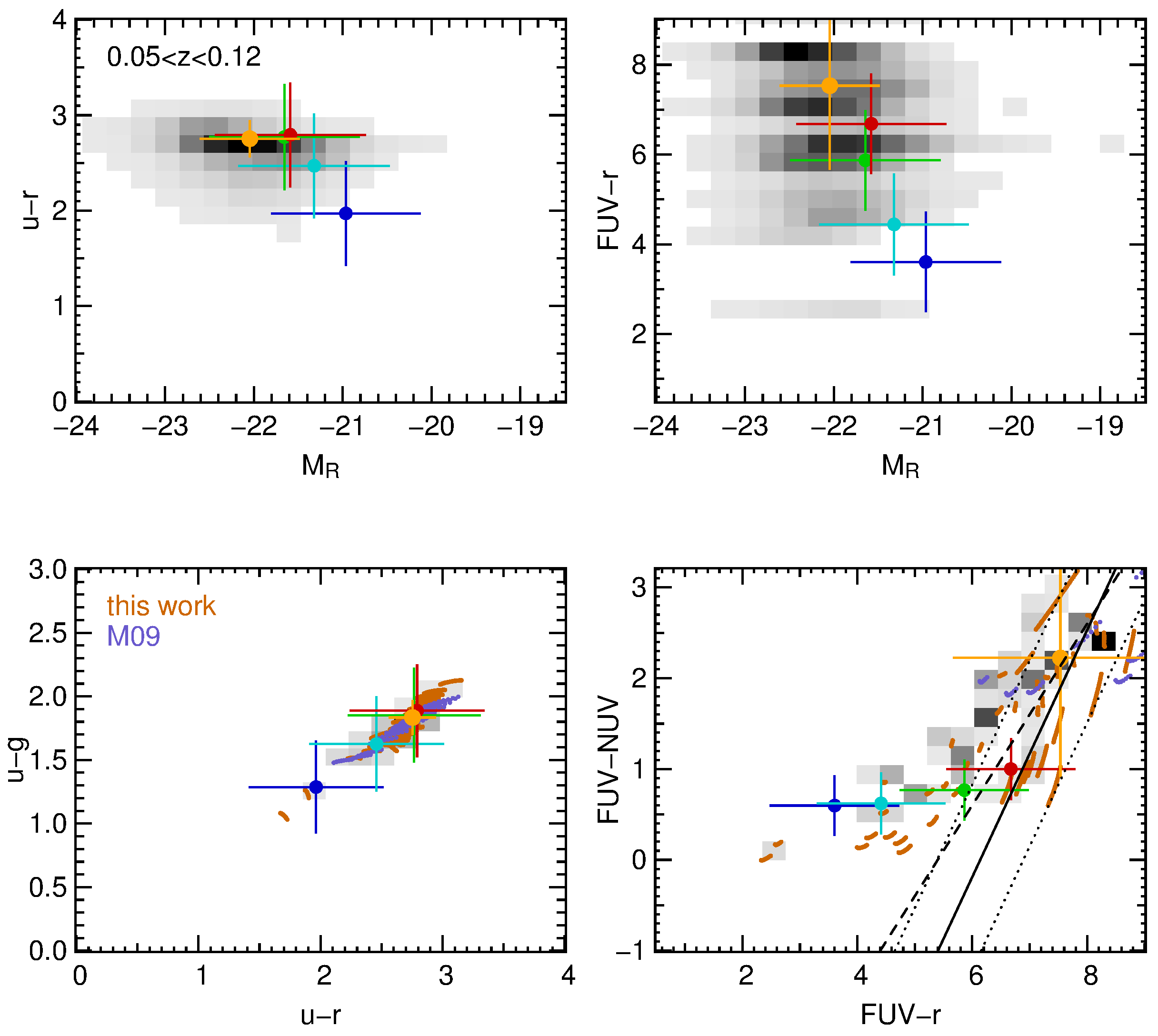}
\caption{UV colors of the new template SEDs from a fit to the spectroscopic redshift of the LRGs within $0.05\leq \zspec\leq0.12$ (gray/black).
	    Yellow filled circles with error bars indicate the median colors and magnitudes with standard deviations.
	    Medians and spreads of R12 are displayed in red (quiescent), green (AGN), cyan (composite) and blue (star-forming ETGs).
	    The orange sequences of dots joining a line segment in the lower two panels indicate the new models from SED fitting, whereas those in purple illustrate the colors of M09 models, both within the same redshift range.
	    The solid line represents the UV red sequence of R12, the dotted lines are $3\times\mathrm{MAD}$.
	    The criterion of RSF galaxies ($\mathrm{NUV}-r<5.4$) found by \cite{2007ApJS..173..512S} is displayed by a dashed line.
	    Plotting ranges correspond to those of R12 (Figure~1).}
\label{fig:UV}
\end{figure*}

We show color\textendash magnitude and the color\textendash color relations of our final template SED set down to UV wavelengths within $0.05\leq \zspec\leq0.12$ and compare our findings to those of \citet[][hereafter R12]{2012ApJ...744L..10R}.
We want to evaluate if the novel models are able to represent the \emph{Galaxy Evolution Explorer} (\emph{\emph{GALEX}}) data, since we used no information about the UV in the creation of our SEDs, as even the \emph{GALEX} NUV filter central rest frame wavelength is $\lambda_{\mathrm{NUV,central}}=2544\,\textrm{\AA}$ at a redshift of $0.12$, whereas the SDSS $u$ band filter has a much larger central wavelength of $\lambda_{\mathrm{u,central}}=3546\,\textrm{\AA}$.\\
R12 selected ETGs in the SDSS data on the basis of morphology, requiring $\mathtt{fracDeV\_r}\approx1$, and removed contamination by late-types through visual inspection.
The sample galaxies were then classified into quiescent, star-forming, composite, and active galactic nucleus (AGN) categories on the basis of their emission line characteristics.
Quiescent ETGs were selected in such a way that they do not exhibit emission lines.\\
Figure~\ref{fig:UV} shows the data medians and their spread of the four ETG categories of the galaxy sample of R12 in color\textendash magnitude ($u-r$ versus $M_R$ and $\mathrm{FUV}-r$ versus $M_R$) and color\textendash color diagrams ($u-r$ versus $u-g$ and $\mathrm{FUV}-r$ versus $\mathrm{FUV}-\mathrm{NUV}$) in analogy to Figure~1 in R12.
Additionally, we plot the colors and magnitudes of our final templates after fitting them to the spectroscopic redshift of the LRGs within $0.05\leq z\leq0.12$, as well as their intrinsic colors within that $z$ range.
The median values of $u-g$ and $u-r$ coincide with those of the quiescent and AGN ETGs of R12 (lower left panel), and we are able reproduce the observed UV-colors.
$\mathrm{med}(M_R)$ on the other hand is about $0.4\,\mathrm{mag}$ lower in this work when compared to the quiescent and AGN ETGs of R12 (upper panels), visualizing that the objects in our sample are definitively LRGs.
Furthermore, the median UV excess ($\mathrm{FUV}-r$) of SEDs of this work is significantly smaller than that of R12 ($\sim1\,\mathrm{mag}$ compared to the quiescent galaxies; upper right panel).
The lower right panel displays the UV-excess versus $\mathrm{FUV}-\mathrm{NUV}$, as well as the UV red sequence found by R12 (solid line) which is derived empirically through a linear fit to the observed colors with width $3\times\mathrm{MAD}$ (dotted lines).
A fraction of our models however lie above the UV red sequence.
Also the $\mathrm{FUV}-\mathrm{NUV}$ color is higher than in R12, due to the fact that our models have higher fluxes in the $\mathrm{NUV}$ band and lower fluxes in $\mathrm{FUV}$.
According to \cite{2007ApJS..173..512S}, recently star forming (RSF) galaxies fulfill the semi-empirical criterion $\mathrm{NUV}-r<5.4$, which is plotted in Figure~\ref{fig:UV} (dashed line).
A considerable part of our models is thus classified as RSF.\\
The lower panels of Figure~\ref{fig:UV} show M09 models on top of the SEDs of this work.
In comparison to R12, M09 models neither cover the region of higher UV-excess ($\mathrm{FUV}-r\lesssim6$), nor those where $\mathrm{FUV}-\mathrm{NUV}\lesssim1.7$, and nearly all of them exhibit no RSF following the criterion of \cite{2007ApJS..173..512S}.
M09 models account for passive evolution only, which is why they do not populate these RSF regions.
However, the M09 SEDs match the $u-r$ versus $u-g$ colors of quiescent, AGN and composite ETGs from R12.
Nevertheless, at low redshifts the LRG sample consists not only of passively evolving galaxies, but shows signs of recent star formation.
M11, which have a large variety of SFHs, ages, and metallicities, should be able to cover these regions.

\subsection{Differences in the SEDs within $z-M_R$ Bins}
\label{sec:zvsmabs}
\begin{figure}[h]
\centering
\includegraphics[width=0.45\textwidth]{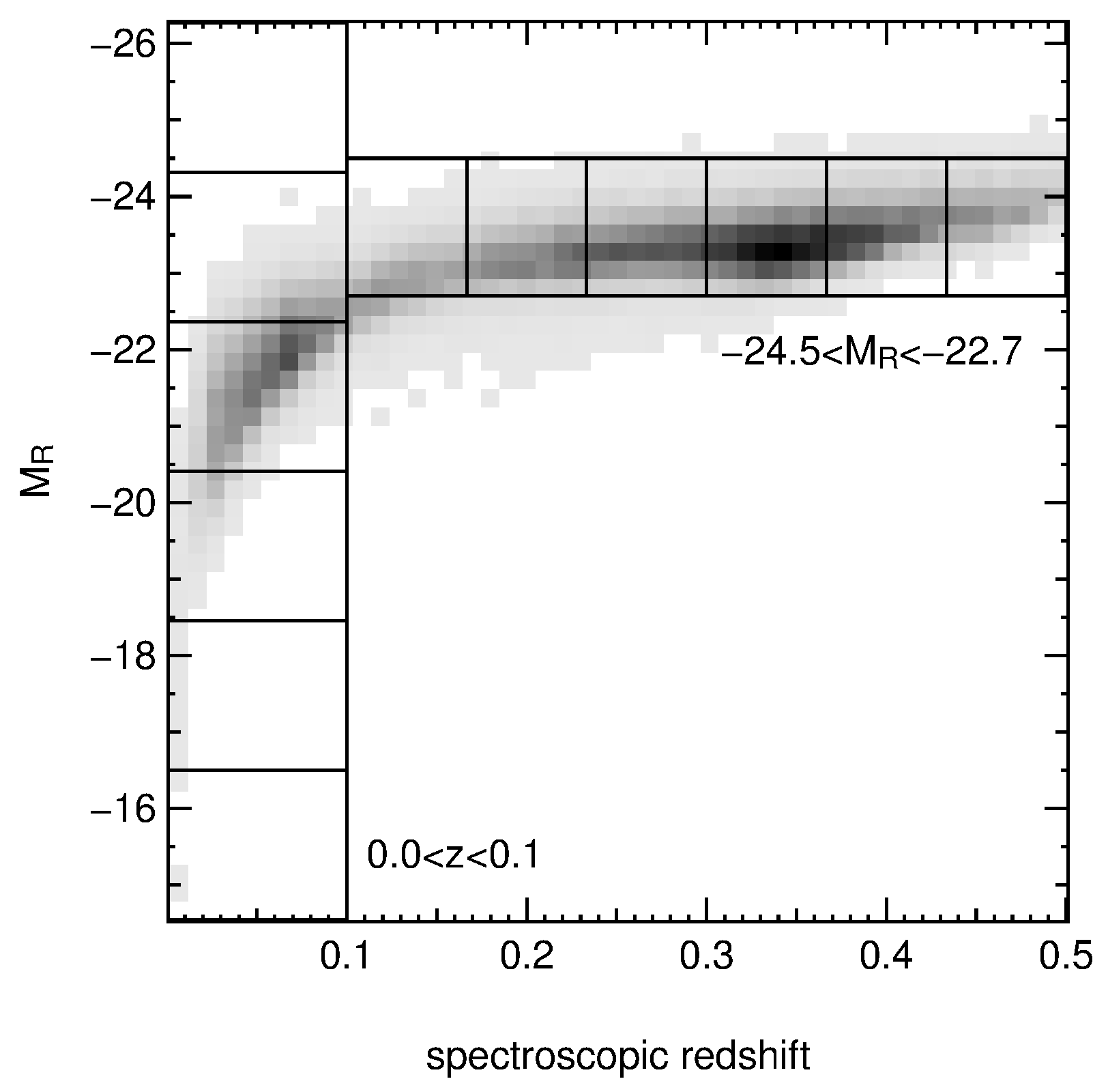}
\caption{
Density plot of spectroscopic $z$ vs. absolute $R$ band magnitude.
$M_R$ is obtained from the template SED which matches a galaxy best at its spectroscopic redshift.
The solid lines show bins within $z$ or $M_R$ intervals.
The luminosity weighted means of the template SEDs that match each object in a given $z-M_R$ bin best are shown in Figures~\ref{fig:ZvsMabs_intM_SED} and \ref{fig:ZvsMabs_intz_SED}.}
\label{fig:ZvsMabs_int3_zvsM}
\end{figure}

\begin{figure}[h]
\centering
\includegraphics[width=0.45\textwidth]{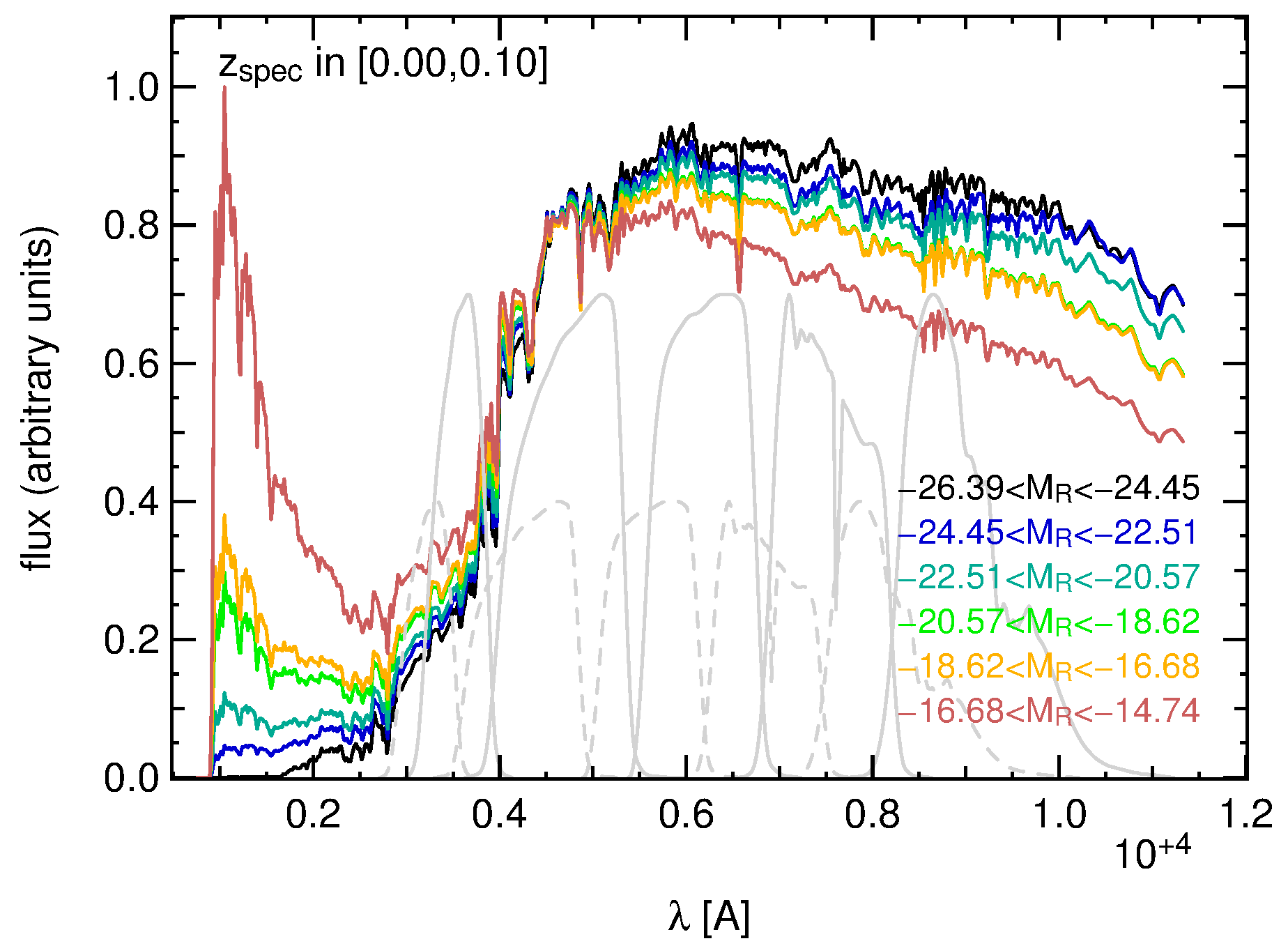}
\caption{Flux weighted superposition of model SEDs (normalized to $f_g$) for objects with $0.0\leq \zspec\leq0.1$ within consecutive luminosity bins.
The solid gray lines are the normalized SDSS filter curves, whereas the dashed gray lines represent the filter curves at $z=0.1$.
The bins contain $1$, $6943$, $23{,}501$, $1757$, $123$, and $33$ objects (from higher to lower luminosities).
}
\label{fig:ZvsMabs_intM_SED}
\end{figure}

\begin{figure}[h]
\centering
\includegraphics[width=0.45\textwidth]{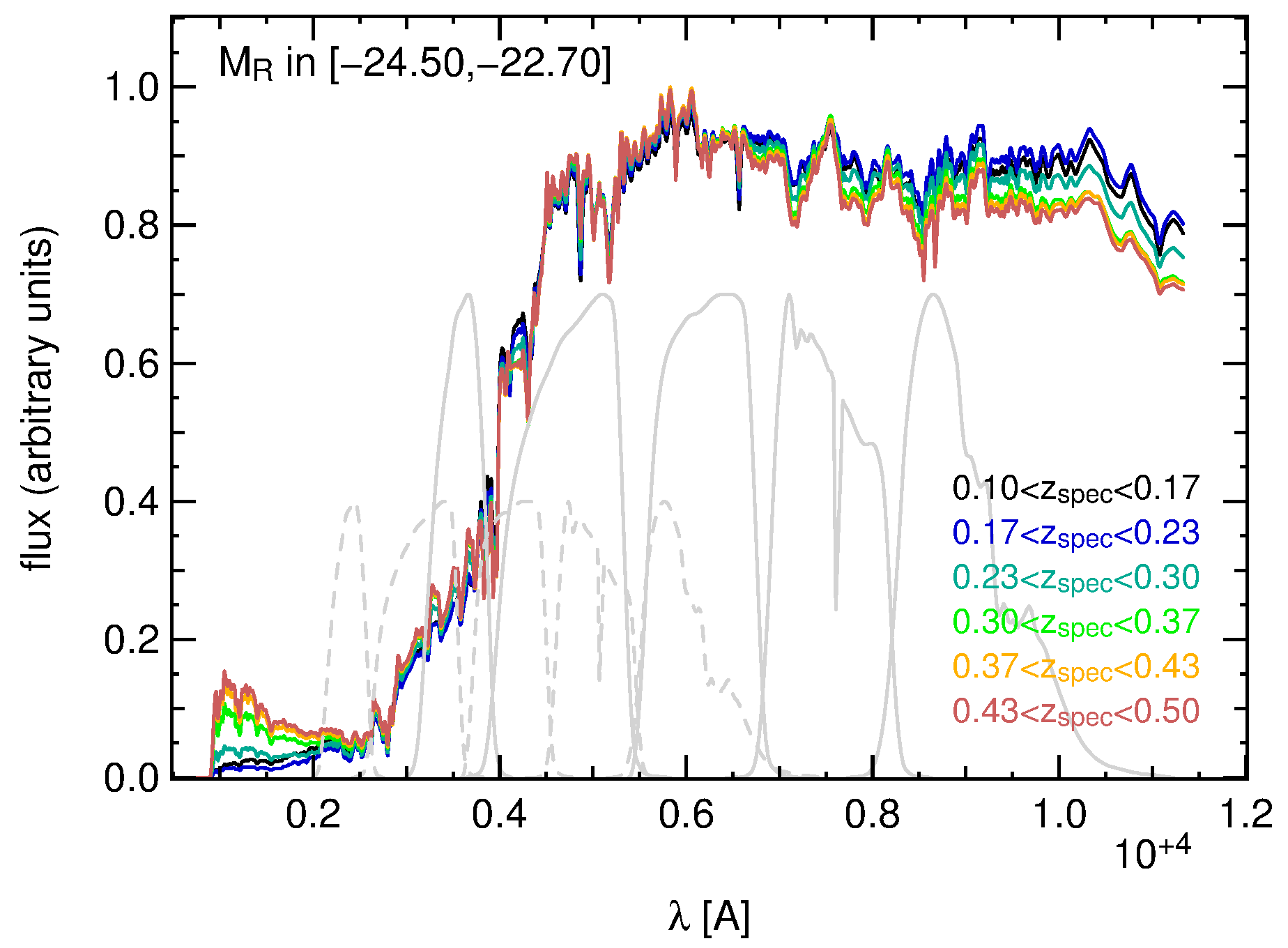}
\caption{
Flux weighted superposition of model SEDs (normalized to $f_g$) for objects with $-24.5\leq M_R\leq-22.7$ within consecutive redshift bins.
The solid gray lines are the normalized SDSS filter curves at $z=0.1$, whereas the dashed gray lines represent the filter curves at $z=0.5$.
The bins contain $7927$, $14{,}750$, $20{,}328$, $30{,}793$, $19{,}648$, and $8702$ objects (from lower to higher redshifts).
}
\label{fig:ZvsMabs_intz_SED}
\end{figure}

By construction our new template SEDs do not necessarily consist only of old CSPs (which are necessary to match the \br\ of the LRGs), but can have imprints of young stellar populations (i.e., more UV-excess relative to old populations) and dust extinction, changing the slope of the CSP and the burst.
Instead of examining what kind of combination the LRGs are fitted by, we present how different the SEDs of galaxies are when matched by our templates.
Therefore, we determine the best fitting SED for every LRG by fitting our new SED set at the spectroscopic redshifts of the objects and calculating the absolute magnitude in the \emph{R} band filter thereafter.
We then bin objects within the $z$ versus $M_R$ plane (Figure~\ref{fig:ZvsMabs_int3_zvsM}) and calculate the mean SED of that bin, by adding up all best fitting SED templates of objects from that bin with weights given by their flux in $R$.
Doing this for different redshift and luminosity bins, we obtain the luminosity weighted light distribution of all objects within a given redshift and luminosity range, under the assumption that they are well described by one of the SEDs from our template set.
The results for luminosity bins with $z\leq0.1$ are shown in Figure~\ref{fig:ZvsMabs_intM_SED}.
The SEDs which match the SDSS data best show a strong dependence on the brightness of the object:
The fainter galaxies are, the more flux they have at $\lambda\lesssim3500\,\textrm{\AA}$ relative to the brightest $M_R\sim-24$ bin, and the steeper the slope of the SED gets for $\lambda\gtrsim5000\,\textrm{\AA}$ on average.
Both observations can be explained by star formation in these galaxies, i.e., that low mass LRGs still build stars at low $z$ in contrast to the more massive LRGs.
Hence, the diversification in the $u-g$ color is not reached by simply selecting smoothly varying SFHs, but rather by adding to these single starburst events of different amounts.
We will confirm this result in Section~\ref{sec:SDSSSF}.\\
Figure~\ref{fig:ZvsMabs_intz_SED} shows the superposition of the models for $z\geq0.1$ and $-24.5\leq M_R\leq-22.7$ in six consecutive redshift bins.
At blue wavelengths ($\lambda\leq3000\,\textrm{\AA}$) the SEDs from objects at higher redshifts exhibit more flux than those at lower redshifts, which indicates increasing star formation turning on to higher $z$.
These effects are in accordance to what one would expect for older stellar populations at low $z$ and younger objects at higher $z$.
Older red populations have (nearly) completed star formation and hence show low fluxes in bluer bands, whereas their fluxes in the red ($\lambda\geq6000\,\textrm{\AA}$) exhibit a smaller continuum slope compared to younger galaxies.
The SED from the first redshift bin ($0.1<\zspec<0.17$) is slightly bluer than the following ($0.17<\zspec<0.23$), because there are still star forming objects within that bin.
This can be seen in Figure~\ref{fig:ZvsMabs_int3_zvsM}: there are still fainter galaxies in this bin, wherefore not all of them are passively evolving.
Additionally, Figure~\ref{fig:ZvsMabs_Ha} in Section~\ref{sec:SDSSSF} confirms that there are still star forming galaxies in this $z$/$M_R$-range.\\
In order to support the latter discussion quantitatively, we performed a $\chi^2$ fit to the variety of BC03 models that are the basis of our template generation in Section~\ref{sec:generation}.
We keep the metallicity fixed at $Z=Z_{\odot}$ as well as $\tau=0.0\,\mathrm{Gyr}$ in order to avoid degeneracies between $Z$, $\tau$, and age.
(We discussed these degeneracies in Section~\ref{sec:BC03colors}.)
The best fitting ages for the redshift\textendash luminosity bins at $z<0.1$ are $9$, $9$, $7$, $4.5$, $4.5$, and $3.5\,\mathrm{Gyr}$ from brighter to fainter bins.
Keeping only the metallicity fixed at $Z=Z_{\odot}$, we observe that the best fitting models nearly all have $\tau=1.0\,\mathrm{Gyr}$, which supports the claim that they all exhibit ongoing star formation.
The first (most luminous) bin has a best fitting $\tau$ of $0.0\,\mathrm{Gyr}$, which is due to the very red SED we calculated for it.
Anyway, since this bin contains only one object we cannot assume the SED to be representative, although it would be in good agreement with the sequence the superpositioned SEDs form, and it would not contradict the observation we make in Section~\ref{sec:SDSSSF} where we deduce from the relations between emission lines that no star-forming galaxies lie in this particular $z-M_R$ range.\\
We also performed the $\chi^2$ fitting of BC03 models to the superpositioned SEDs for $z>0.1$ and $-24.5\leq M_R\leq-22.7$ keeping $Z$ at $Z_{\odot}$ and $\tau=0.0\,\mathrm{Gyr}$ fixed.
The best fitting ages are then: $12$, $13$, $11$, $9$, $9$, and $9\,\mathrm{Gyr}$ with increasing $z$.
As was mentioned above, the first redshift bin ($0.1<\zspec<0.17$) still contains objects with ongoing star formation, which is why the best fitting age is lower than in the second bin ($0.17<\zspec<0.23$).
Nevertheless, the ages from the other five redshift slices are monotonically decreasing.
By keeping only the metallicity fixed, the $\tau$ values increase from lower to higher redshift, which also yields to bluer spectra.
Furthermore, the differences in the light travel time between the centers of the first and the last redshift bins is $\sim3.1\,\mathrm{Gyr}$, which is in agreement with the $\chi^2$ fit results, and also supports what we expect when we assume that the underlying population of our sample at $z>0.1$ predominantly consists of passively evolving galaxies, and that the variations in the SEDs are a result of aging.

\subsubsection{Comparison to SDSS Star Forming LRGs}
\label{sec:SDSSSF}

\begin{figure}[h]
\centering
\includegraphics[width=0.4\textwidth]{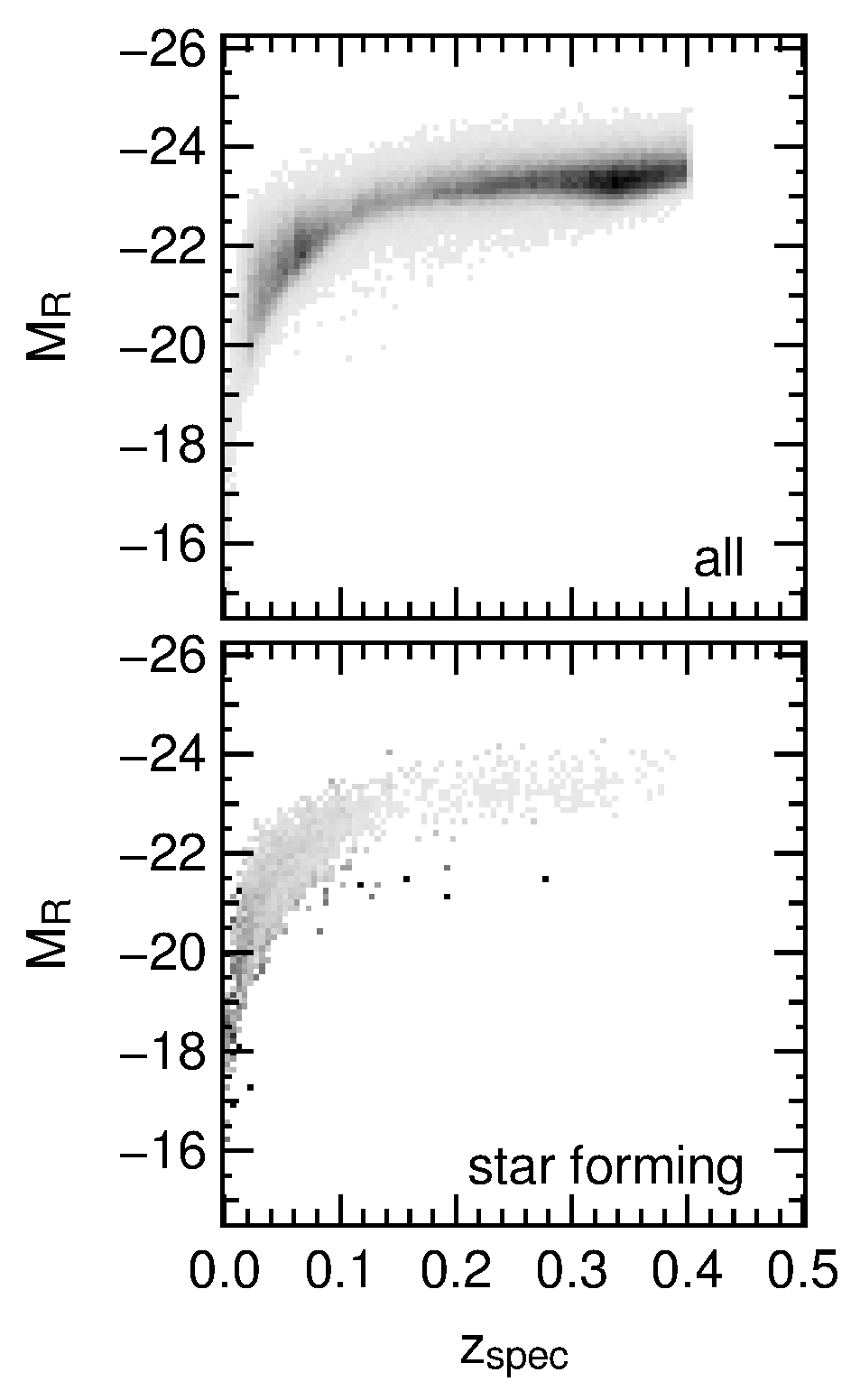}
\caption{
Density plots of $z$ vs. absolute $R$ band magnitude of LRGs.
Upper panel: all LRGs where $H\alpha$ lies within the wavelength coverage of SDSS.
Lower panel: Relative abundance of LRGs fulfilling $\log_{10}(\mathrm{OIII}/\mathrm{H\beta}) < 0.7 - 1.2(\log_{10}(\mathrm{NII}/\mathrm{H\alpha}) - 0.4)$ and hence showing star formation activity.
}
\label{fig:ZvsMabs_Ha}
\end{figure}

\begin{figure}[h]
\centering
\includegraphics[width=0.45\textwidth]{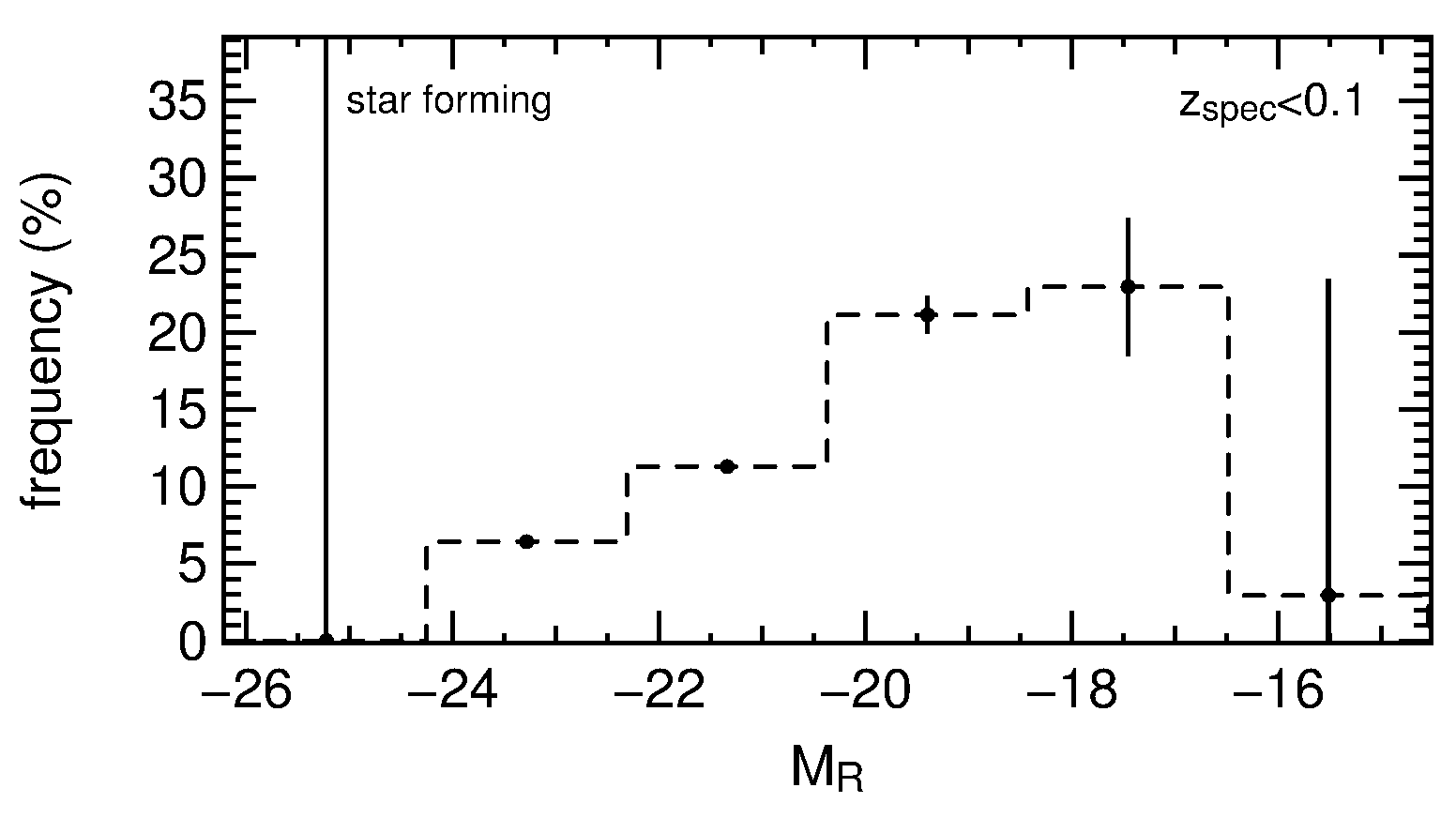}
\caption{
Percental fraction of $z<0.1$ star forming LRGs as a function of luminosity. Bins are the same as in Figure~\ref{fig:ZvsMabs_int3_zvsM} and \ref{fig:ZvsMabs_intM_SED}. For a definition of ``star forming'' see the text. Error bars indicate the Poissonian errors for the intermediate bins, and Binomial errors (with $84\%$ confidence) for the outer bins which are not as highly populated. The bins contain (from higher to lower luminosity) $1$, $6917$, $23{,}429$, $1751$, $122$, and $34$ objects in total (star-forming and non-star-forming).
}
\label{fig:ZvsMabsbins_Hahist}
\end{figure}

\begin{figure}[h]
\centering
\includegraphics[width=0.45\textwidth]{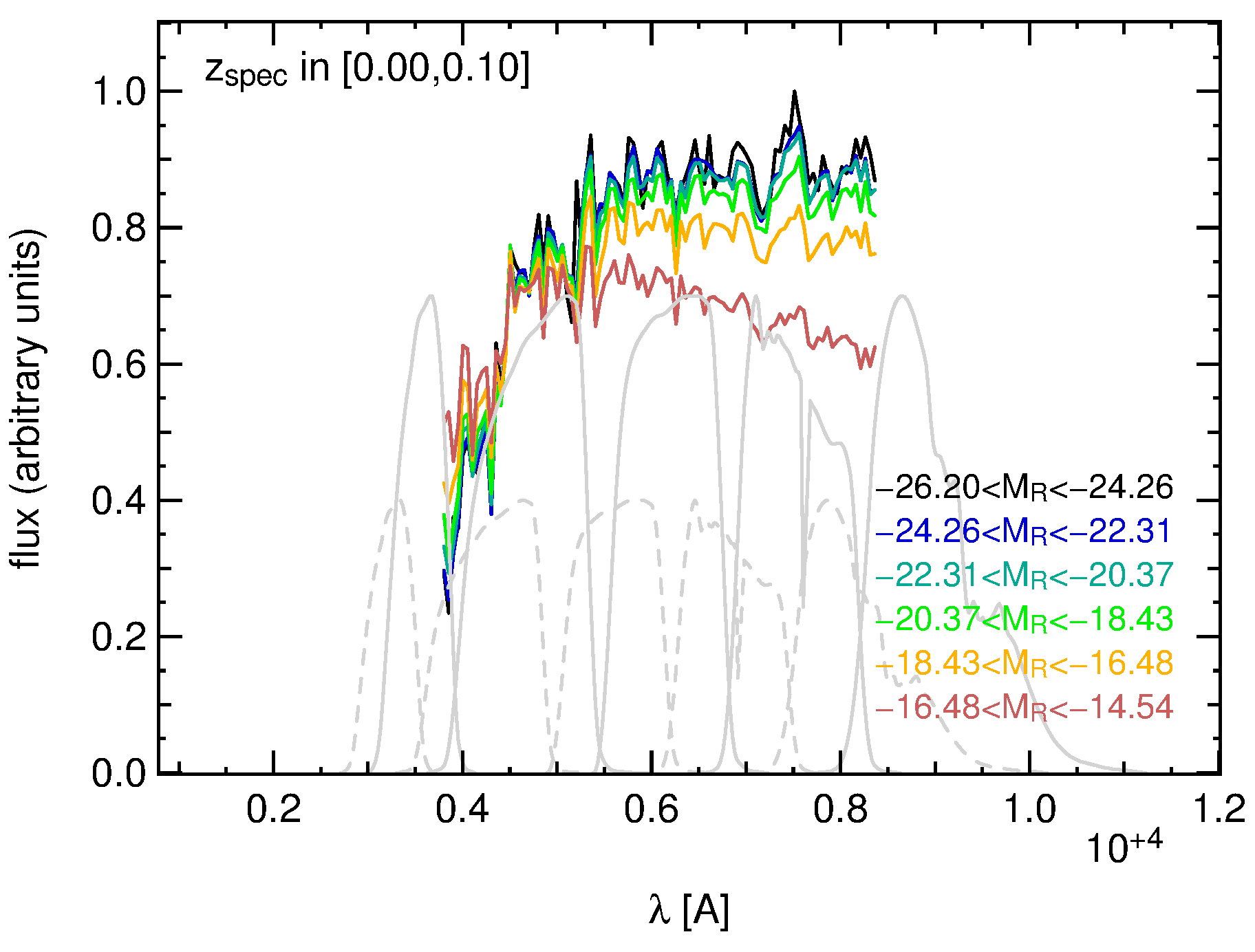}
\caption{
Stacked rest frame spectra for SDSS LRGs at $z<0.1$ and within the same $z-M_R$ bins as in Figures~\ref{fig:ZvsMabs_int3_zvsM}, \ref{fig:ZvsMabs_intM_SED}, and \ref{fig:ZvsMabsbins_Hahist}.
The SEDs were normalized to $f_g$, interpolated to a resolution of $50\,\textrm{\AA}$, and weighted by the calculated $f_R$ (likewise as those in Figure~\ref{fig:ZvsMabs_intM_SED}) in the stacking.
Different colors represent different $M_R$ bins.
The spectra are shown within the combined SDSS wavelength coverage for objects between $z=0.0$ and $0.1$, whereas plotting limits are the same as in Figure~\ref{fig:ZvsMabs_intM_SED}.
SDSS filter transmissions are displayed by solid ($z=0.0$) and dashed ($z=0.1$) gray lines.
}
\label{fig:specstack}
\end{figure}

We want to compare our findings of the previous chapter to the observed spectra of SDSS LRGs.
Therefore, we download the SDSS spectral subclassifications of our LRG sample that were produced by the \texttt{SpecBS} pipeline, where galaxies are further categorized by their emission lines.
Galaxies that fulfill $\log_{10}(\mathrm{OIII}/\mathrm{H\beta}) < 0.7 - 1.2(\log_{10}(\mathrm{NII}/\mathrm{H\alpha}) - 0.4)$ are tagged as ``star forming''.
If they furthermore have an $H\alpha$ equivalent width of equal or greater than $500\,\textrm{\AA}$, they are classified as ``starburst''.
In our LRG sample only $10$ objects fulfill the ``starburst'' criteria.
In the following, we will refer to both tags (``starforming'' and ``starburst'') as \emph{star forming}.
Since the spectral wavelength coverage of SDSS is $[3800\,\textrm{\AA},9200\,\textrm{\AA}]$, the $H\alpha$ line drops out of the sample at $z>0.4$.
From the previously investigated $\sim140{,}000$ objects, $\sim120{,}000$ remain in the sample when we cut off LRGs that have $z>0.4$.
From these, $4000$ show indications of star formation.
The upper panel of Figure~\ref{fig:ZvsMabs_Ha} shows the density of LRGs in the $z-M_R$ plane, and the lower panel displays the relative abundances of LRGs exhibiting star formation.
Here, one can observe that the fraction of local star forming LRGs increases toward smaller luminosities.
Additionally, we show in Figure~\ref{fig:ZvsMabsbins_Hahist} the percental fractions of star-forming LRGs for the same redshift-absolute $R$ band magnitude bins that we investigated in Section~\ref{sec:zvsmabs}, Figures~\ref{fig:ZvsMabs_int3_zvsM} and \ref{fig:ZvsMabs_intM_SED}.
The errors in the intermediate bins are Poissonian errors.
The errors of the outer two luminosity bins, which do not contain as many objects, are Binomial with a confidence level of $84\%$.
In principle, the highest and lowest $M_R$ bins could be neglected, since they contain only $1$ and $33$ objects, respectively.
Except for the region of faint galaxies (the lowest luminosity bin), the frequencies of star-forming LRGs shows an increase toward fainter magnitudes.
This is in agreement with what we would expect from Figure~\ref{fig:ZvsMabs_intM_SED}, where we saw that the mean model SEDs for faint LRGs are bluer and show more star-forming activity, as well as with the results of the $\chi^2$ fit we performed in the last section.\\
In addition to that we downloaded spectra for random subsamples of the objects with $z<0.1$ and stacked them.
Thereby we weighted them by the best fit SED flux in the $r$ band as we did with the superposition of our models in Section~\ref{sec:zvsmabs}.
In each luminosity bin the spectra of a subsample containing $100$ objects (less, if the bin contains less objects) were superpositioned.
We show the results in Figure~\ref{fig:specstack}, where we consider the added spectra in the rest frame wavelength range which is covered by SDSS for objects with $z<0.1$.
Comparing Figure~\ref{fig:specstack} to Figure~\ref{fig:ZvsMabs_intM_SED}, we observe that the stacked SDSS spectra show a very similar behavior as the average best fitting SEDs that were inspected previously, supporting that our novel SEDs are able to describe the data of this population.\\
At greater redshifts ($z\gtrsim0.1$), no faint galaxies are observable, and only $370$ ($\sim0.5\,\%$) LRGs are classified as starforming.
This region is therefore strongly dominated by objects that show no recent star formation.
The variations we saw in Figure~\ref{fig:ZvsMabs_intz_SED} are not caused by different fractions of star forming galaxies (which are anyhow of very low abundance), but result from aging only, which was also supported by the $\chi^2$ fit to BC03 models we performed in Section\ref{sec:zvsmabs}.\\
We also stacked spectra of objects from these $z-M_R$ bins in the same manner as above.
Unfortunately, one cannot see the effect of a flatter NIR and declining UV flux at low $z$ compared to a declining NIR and increasing UV flux at high $z$ as clearly as in Figure~\ref{fig:ZvsMabs_intz_SED}, because the NIR part is only observable for objects with low $z$, and the UV only for objects with high $z$.
However, the trends at low $z$ (i.e., flat NIR flux) and at high $z$ (increasing UV flux) are qualitatively there.

\section{Summary}
\label{sec:summary}
In this work we analyzed the colors as a function of redshift of a number of publicly available red model SEDs in the SDSS filter system and compared them to the colors of spectroscopic SDSS LRGs at redshifts from $z=0$ to $0.5$.
We found that no investigated model set can describe the $u-g$, $g-r$, $r-i$, and $i-z$ colors of the data at all redshifts.
Thus we created new red model SEDs for five redshift bins on the basis of the LRG photometry by fitting a superposition of a BC03 CSP model and a burst SED, while allowing for dust extinction of both components.
We estimated photometric redshifts with the Bayesian template fitting code \photoz\ and a template set containing the created SEDs with according redshift and luminosity priors.
The resulting catastrophic outlier rate is $\eta=0.12\%$, whereas the scatter reads $\sigma_{\dz}=0.027$ and $\snmad=0.017$ for non-outliers.
The overall bias reaches $\langle|\dz|/(1+\zspec)\rangle=0.015$.
We compared our results with photometric redshift estimates available in the SDSS database, both for the template fitting approach and for the neural networks.
Our $\langle|\dz|/(1+\zspec)\rangle$, $\sigma_{\dz}$, and $\snmad$ are of the same order as the SDSS database template fitting values, although with a slightly higher catastrophic outlier rate.
Nevertheless, the medium bias varies significantly less from $\dz=0$ with our template SEDs.
The total values of our scatter, bias and outlier rates marginally differ from the SDSS ANN(CC2) photometric redshifts and are worse than the results of ANN(D1).
However, the bias of ANN redshifts reaches higher values within some redshift ranges.\\
We investigated the properties of the novel SEDs and demonstrated that they also describe the variety of \emph{GALEX} NUV and FUV colors well.
Performing an SED fit at the spectroscopic redshift of the LRGs, we observed that the flux weighted mean SEDs of local galaxies ($z\leq0.1$) vary notably as a function of luminosity, meaning that faint galaxies show increasing indications of ongoing star formation.
The stacked spectra from the SDSS show a very similar behavior.
Furthermore, we find that the fraction of star forming SDSS LRGs at $z<0.1$ increases with decreasing $R$ band luminosity.
This is in agreement with our finding that the mean SED is bluer for faint local galaxies.
We also observe a difference in the flux weighted mean SEDs as a function of $z$, which is an effect of aging only and not due to recent star formation.
In Appendices~\ref{app:models} and \ref{app:colcol} we showed that BC03 models cover the data better than M09 and M11 in color space, which is why we favored the BC03 relative to the M09 and M11 model sets for the creation of our new SEDs.
Finally, we also showed how our new template SED colors compare to the data in the $ugri$, $griz$, $grri$, and $riiz$ planes.\\
We would be happy to provide the photometric redshifts together with the rest frame magnitudes for the complete photometric LRG sample of SDSS on request.\\

\acknowledgments
We thank Ulrich Hopp for helpful suggestions concerning the spectral analysis, and Ben Hoyle for the acquisition of the SDSS spectra.
We are very grateful to the anonymous referee for her/his suggestions that greatly improved the manuscript.
Furthermore, we want to thank Claudia Maraston for reading and commenting an early version of this manuscript.\\
This work was supported by SFB-Transregio 33 (TR33) ``The Dark Universe'' by the Deutsche Forschungsgemeinschaft (DFG) and the DFG cluster of excellence ``Origin and Structure of the Universe``.

\bibliographystyle{apj}
\bibliography{apj-jour,bibliography}

\appendix
\section{SED Fitting Results with Different Model Sets}
\label{app:models}
\begin{figure}
\centering
\leavevmode
\columnwidth=.45\textwidth
\vspace{0.2cm}
\includegraphics[width={\columnwidth}]{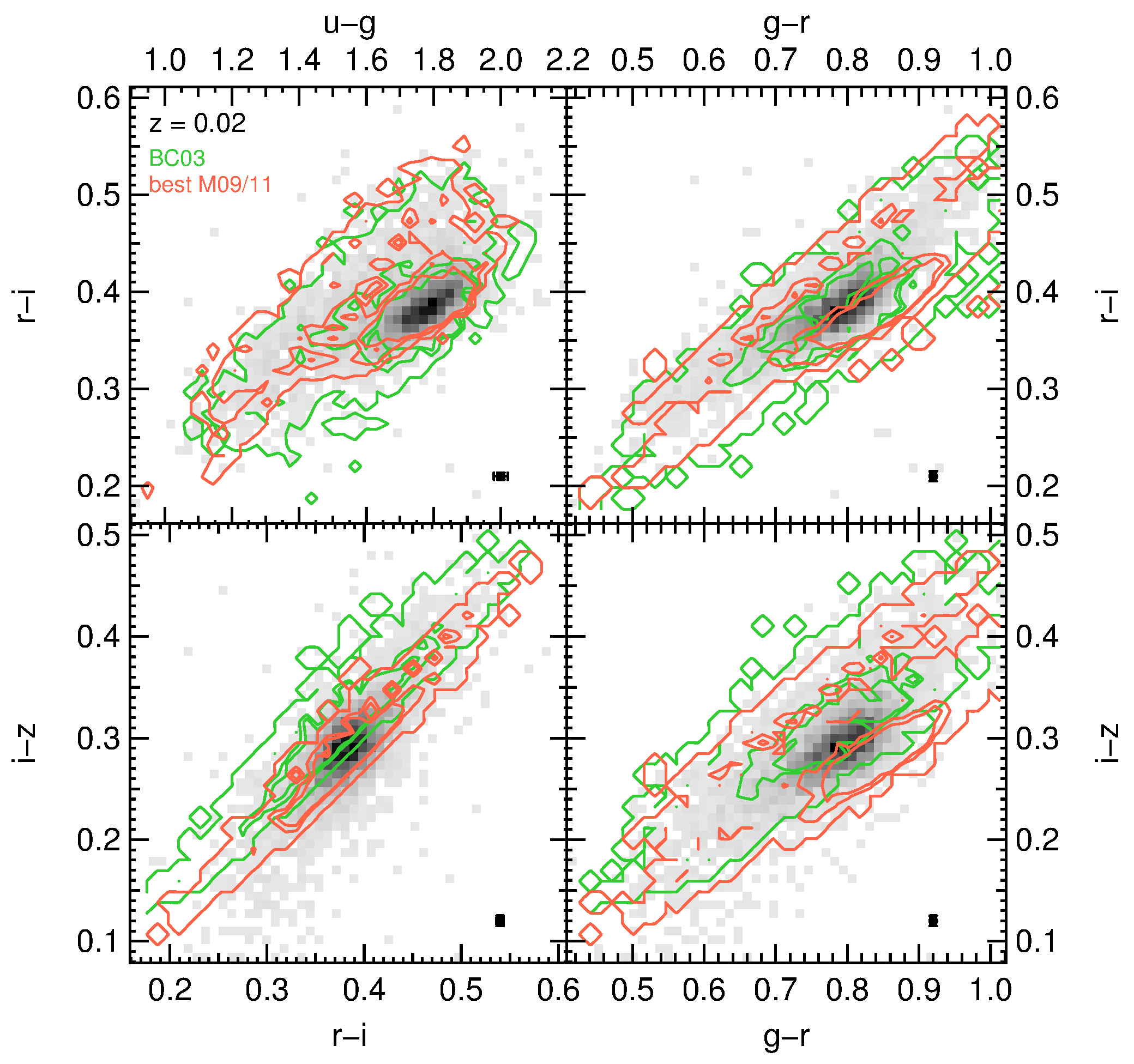}
\hfil
\includegraphics[width={\columnwidth}]{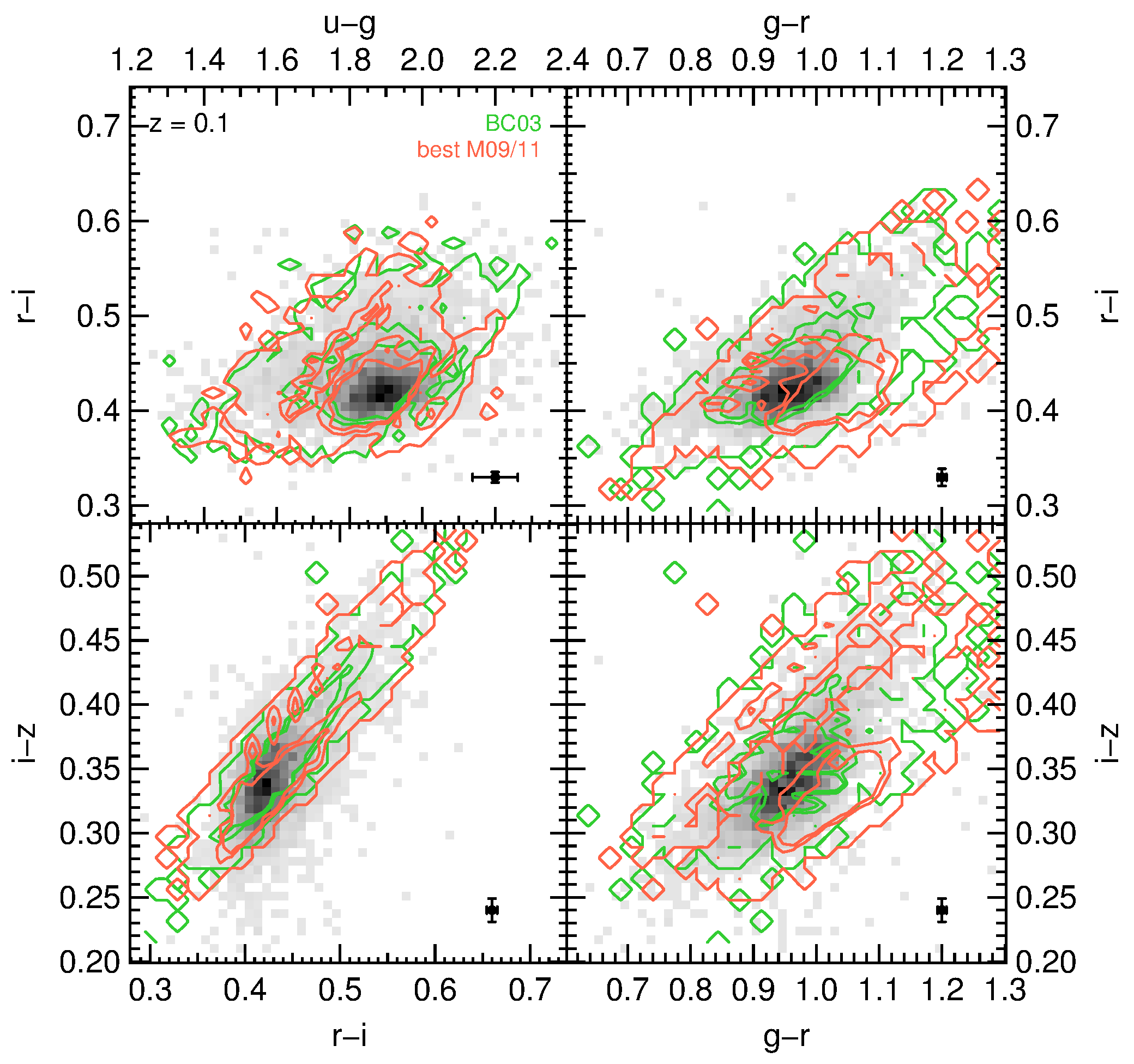}
\vspace{0.2cm}
\includegraphics[width={\columnwidth}]{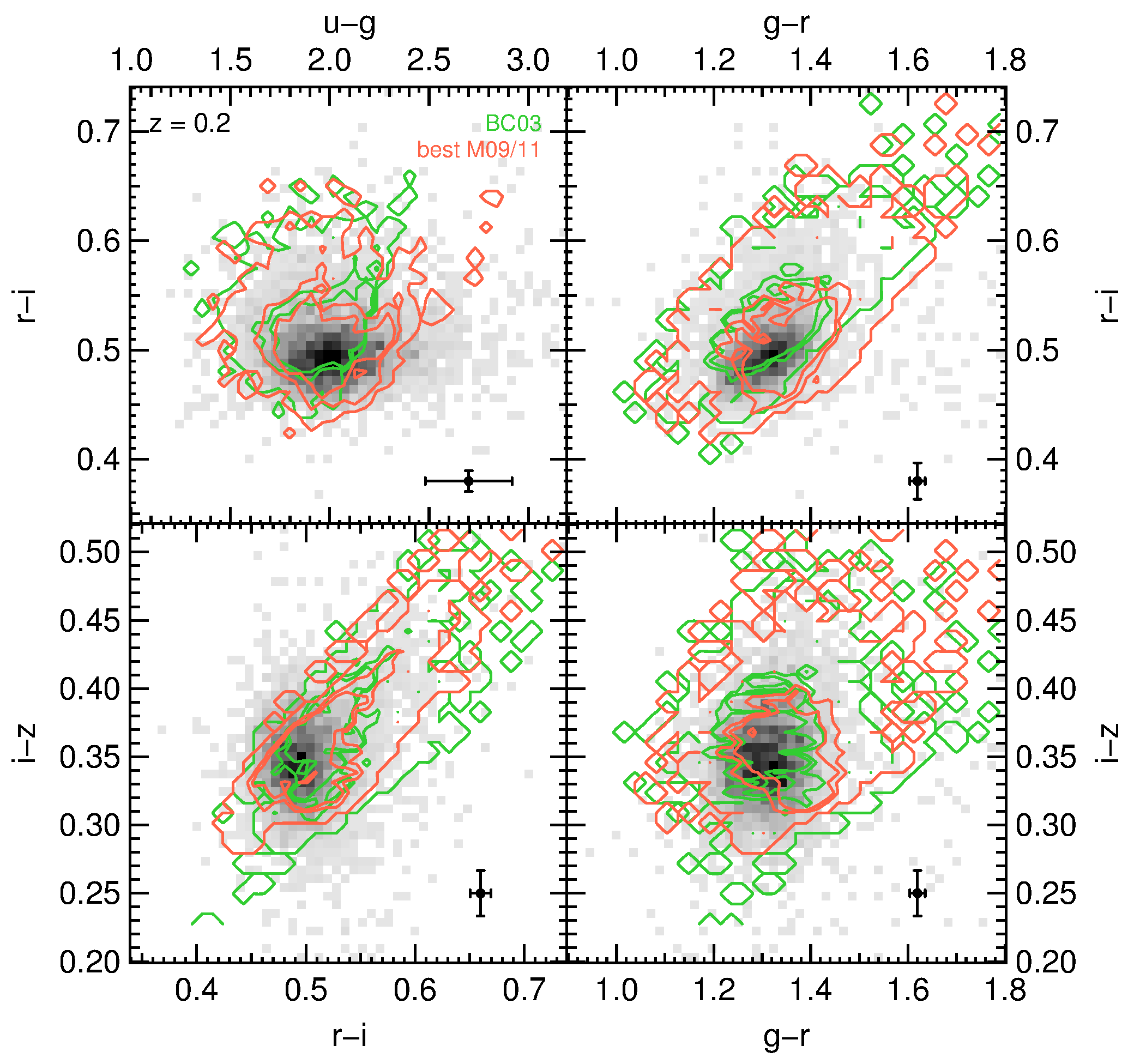}
\hfil
\includegraphics[width={\columnwidth}]{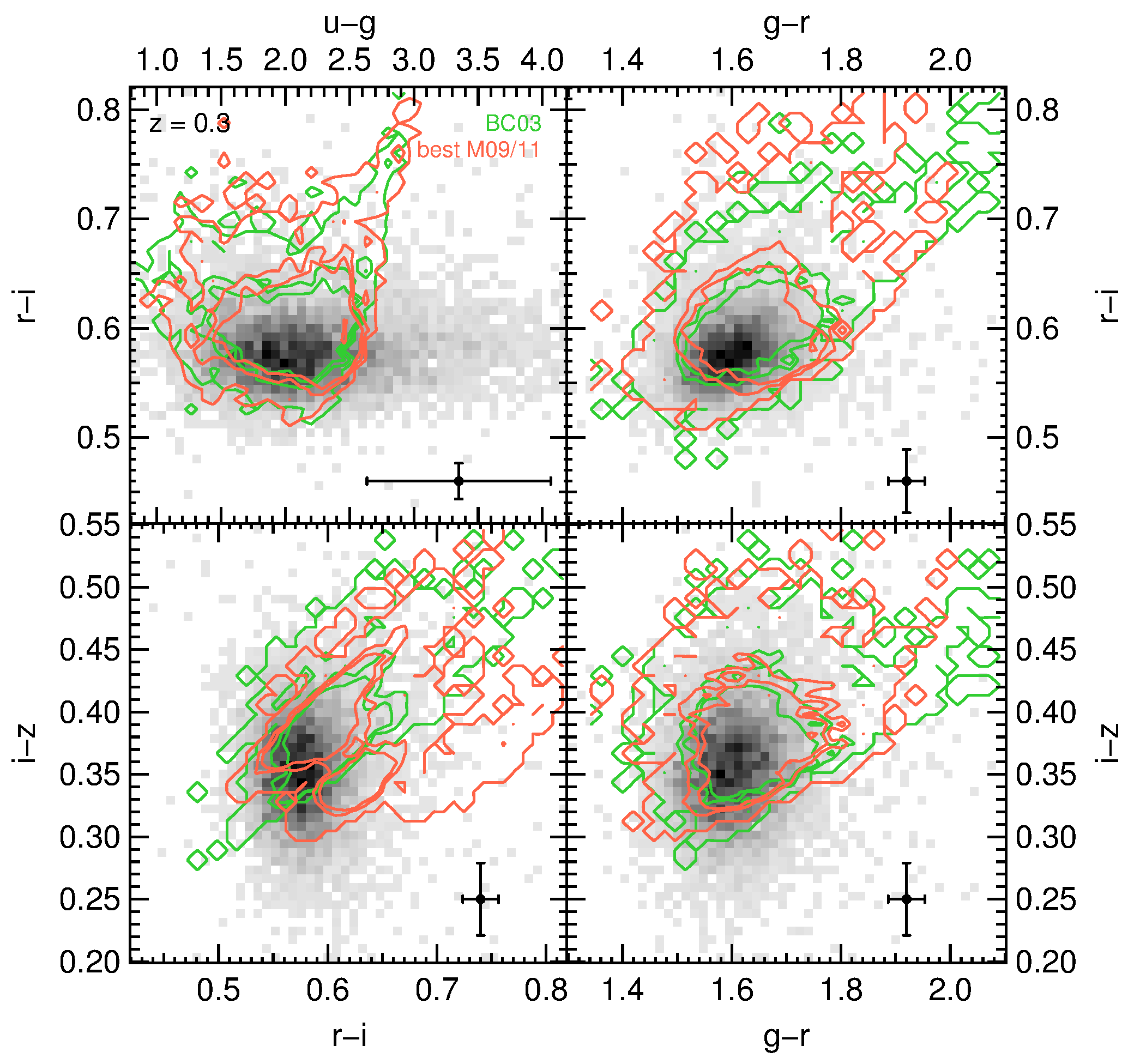}
\includegraphics[width={\columnwidth}]{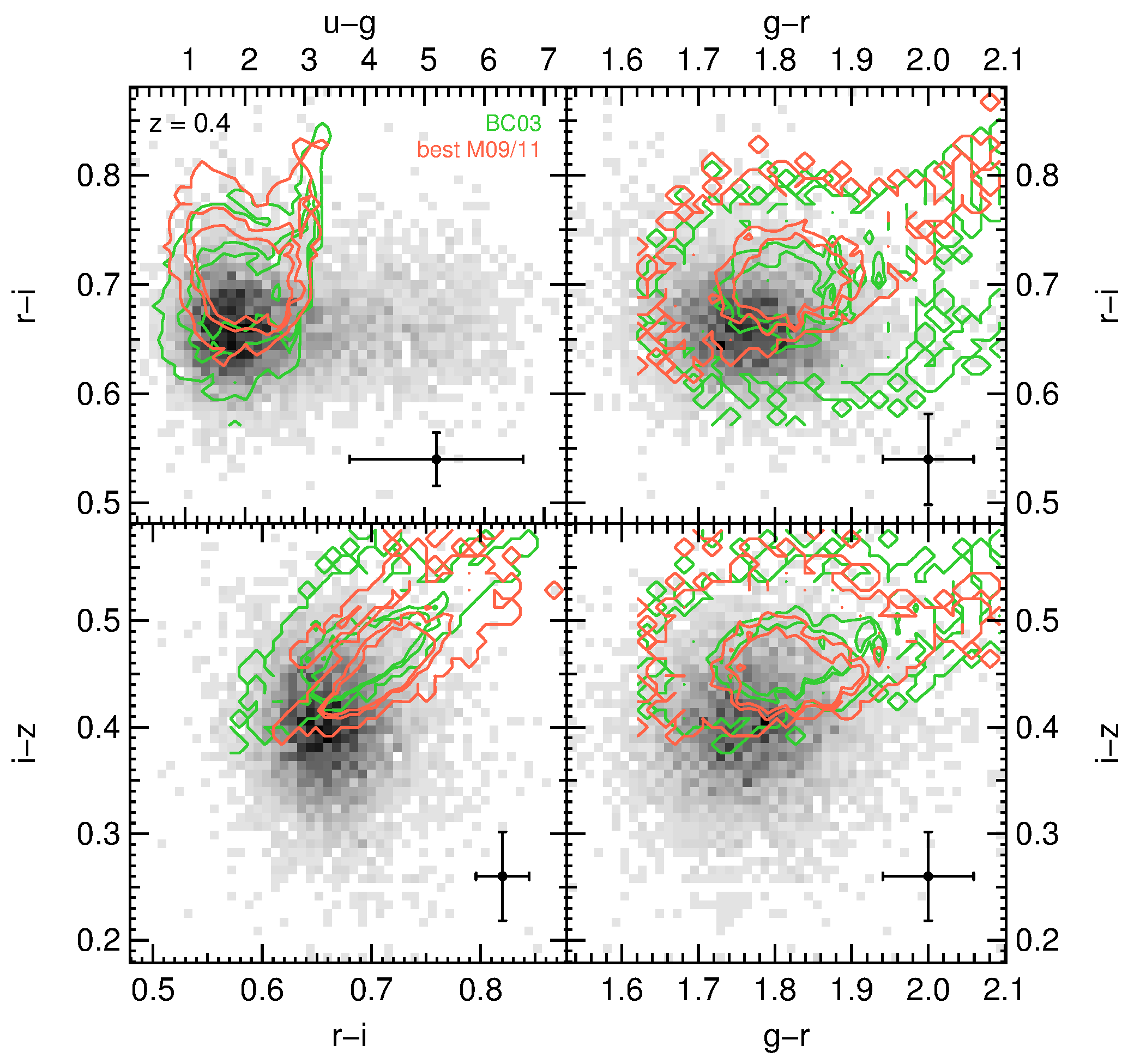}
\hfil
\caption{
Color distribution of LRGs (gray). Contours show the SED fitted BC03 (green), and M09/11 (red) models superposed with a burst model to fit the LRGs.
}
\label{fig:COLSEDfit}
\end{figure}

\begin{figure}
\centering
\leavevmode
\columnwidth=.4\textwidth
\vspace*{0.4cm}
\includegraphics[width={\columnwidth}]{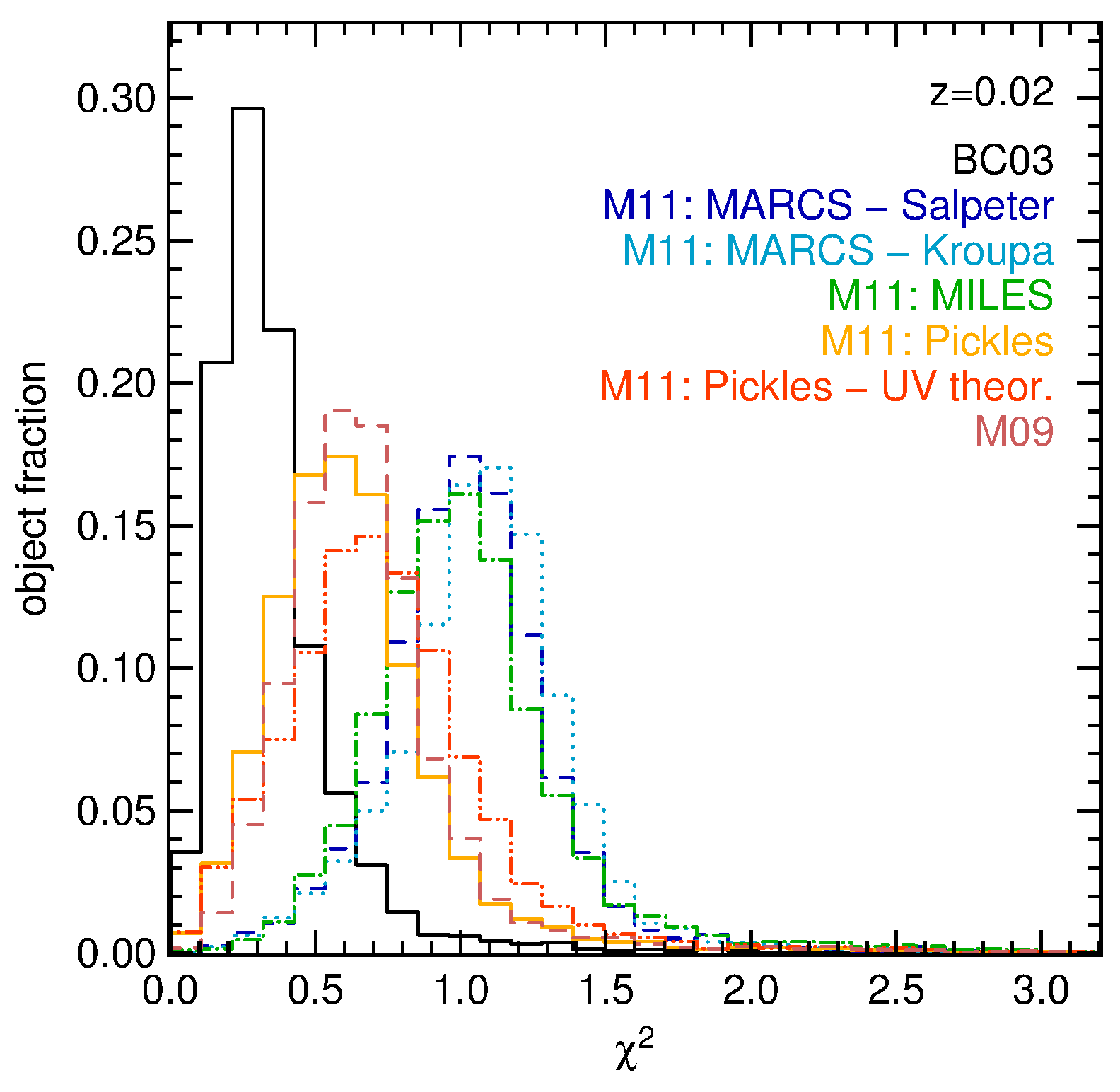}
\hfil
\includegraphics[width={\columnwidth}]{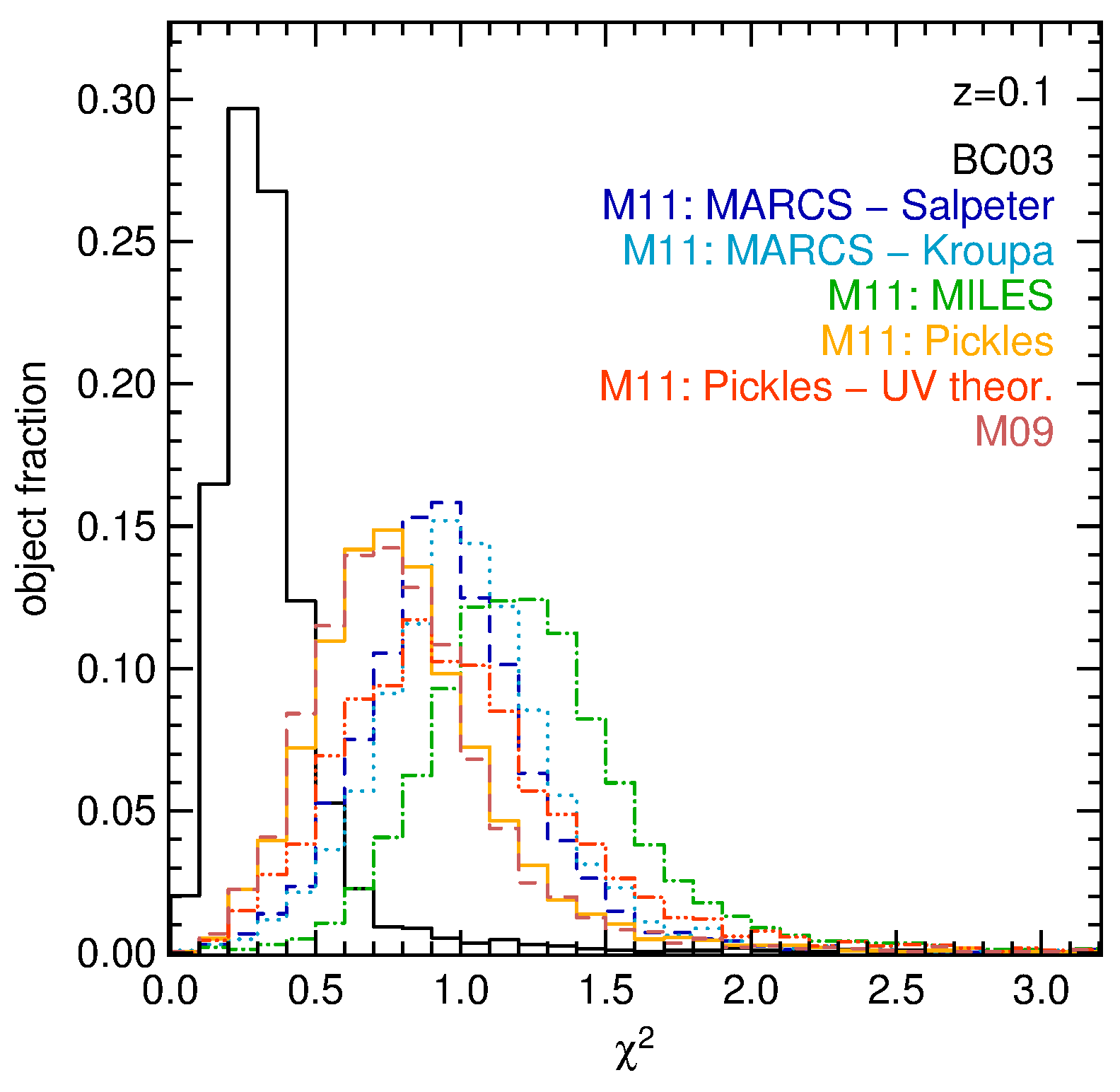}
\vspace*{0.4cm}
\includegraphics[width={\columnwidth}]{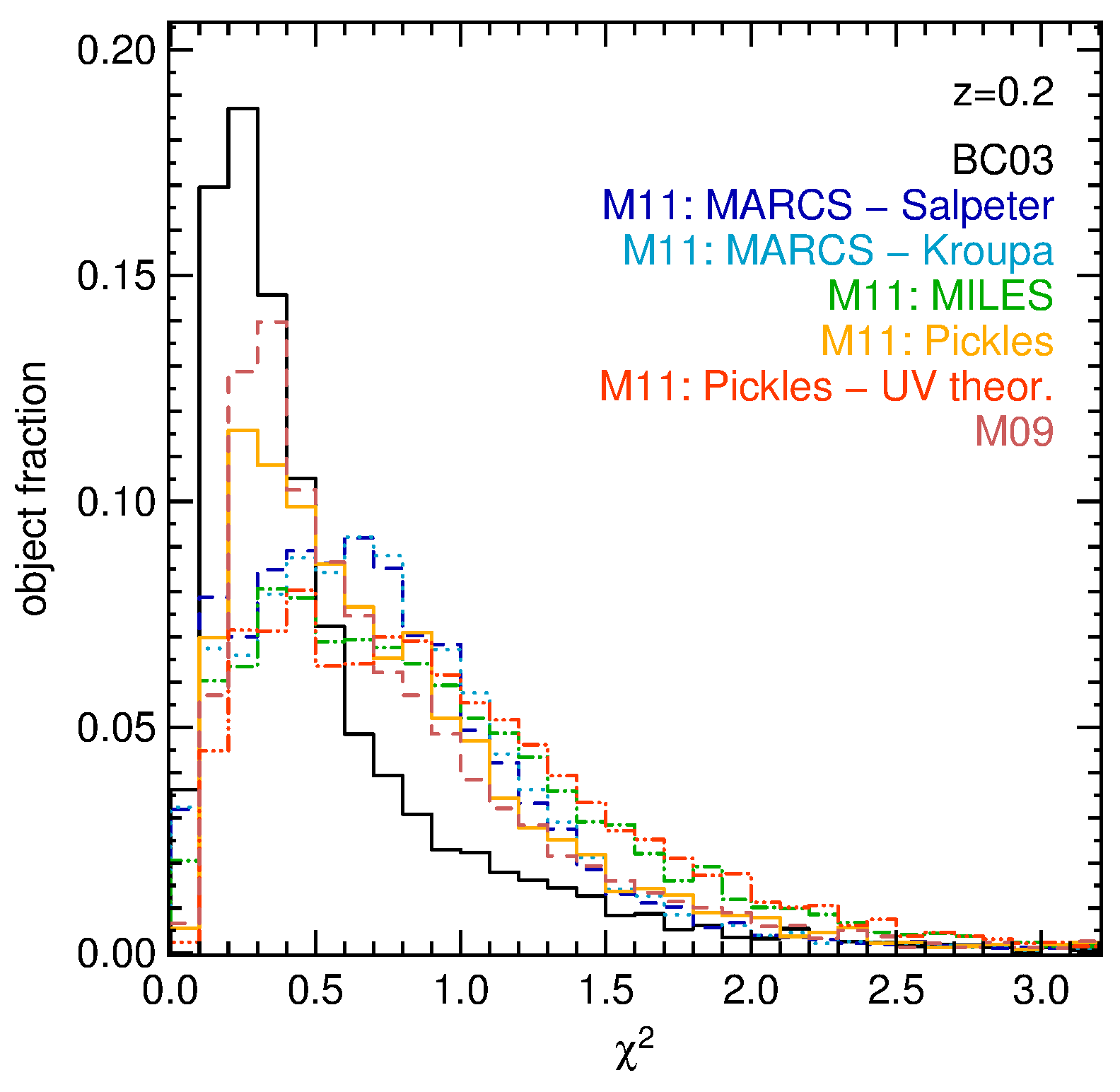}
\hfil
\includegraphics[width={\columnwidth}]{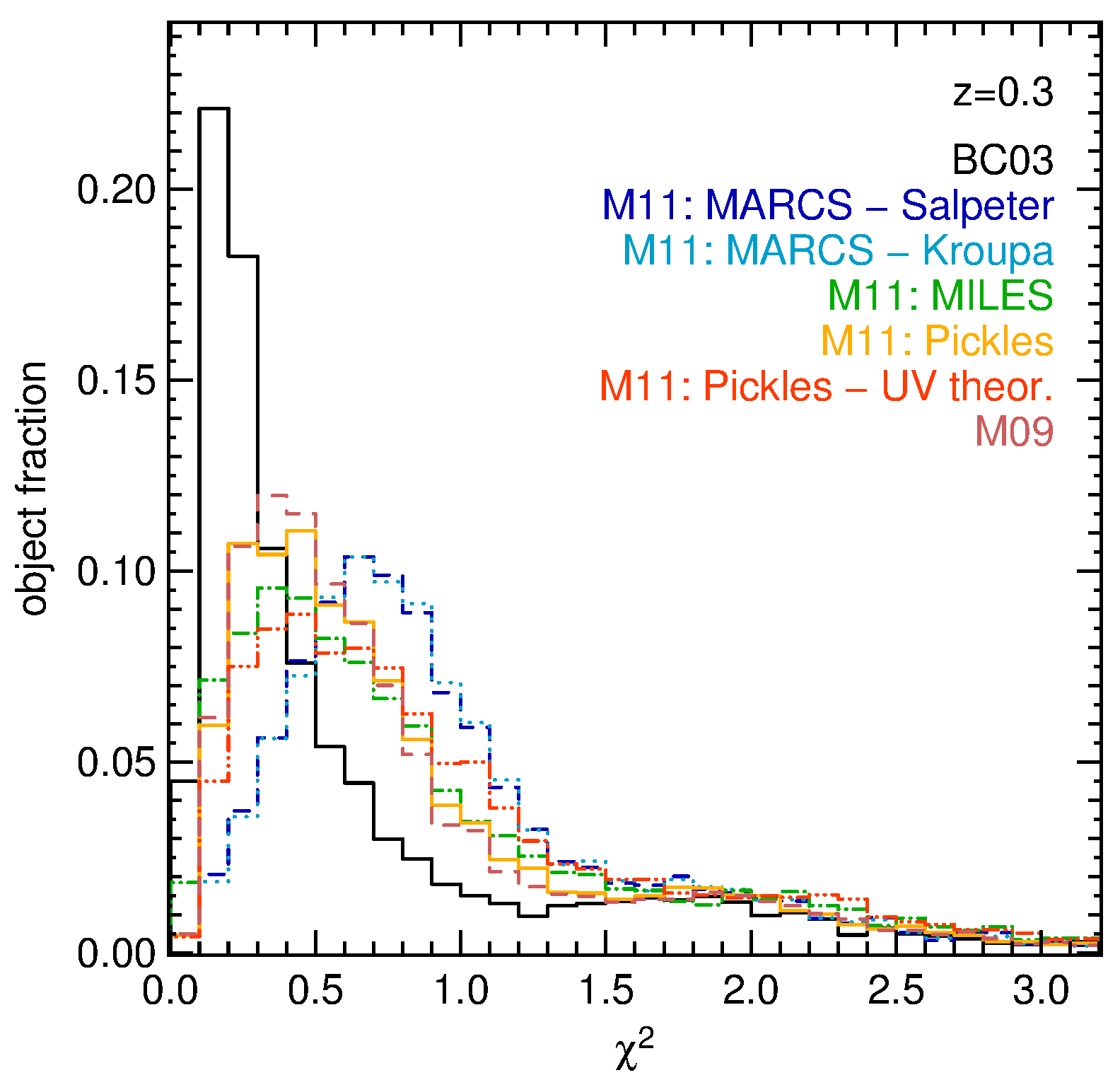}
\includegraphics[width={\columnwidth}]{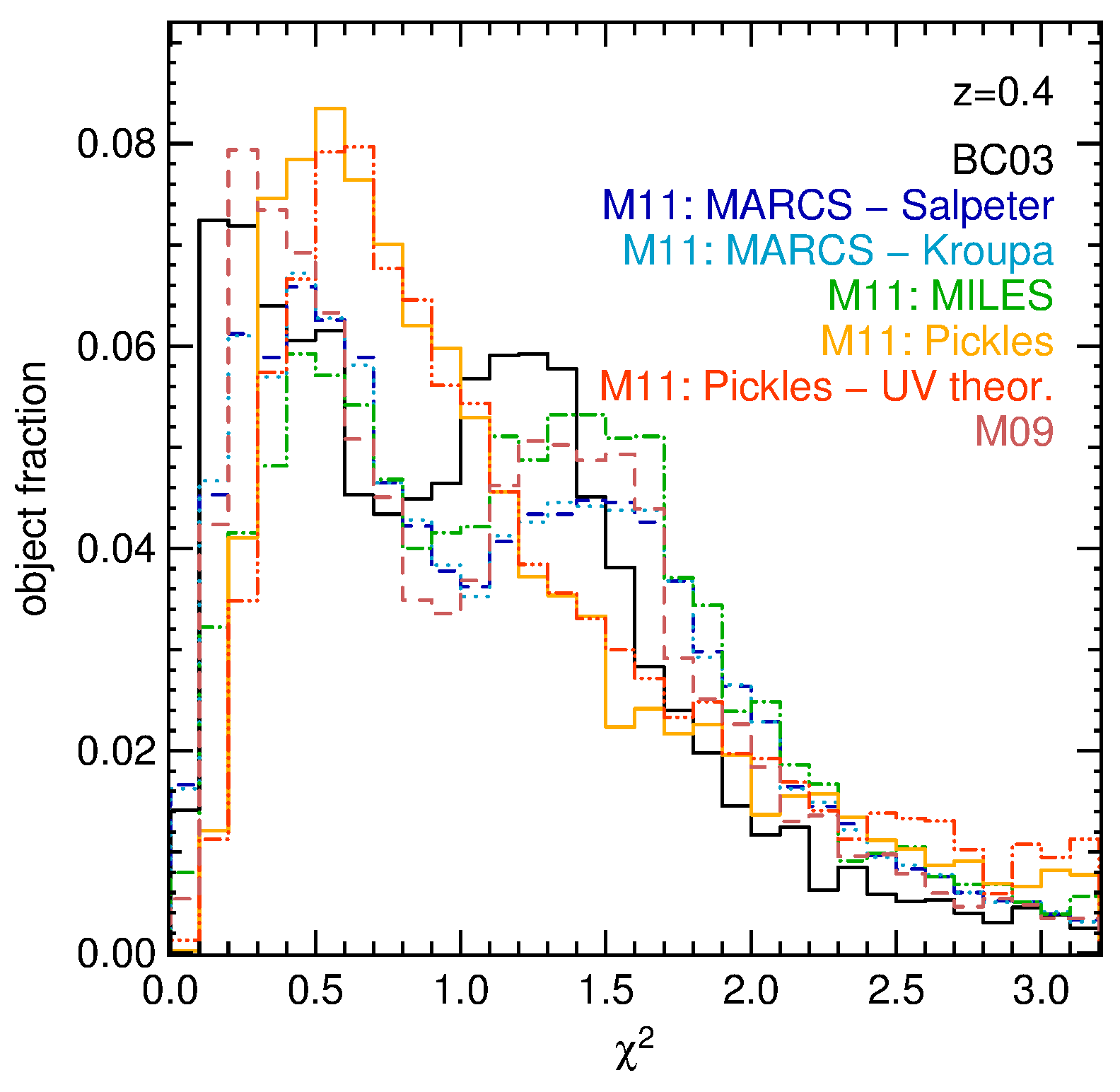}
\hfil
\includegraphics[width={\columnwidth}]{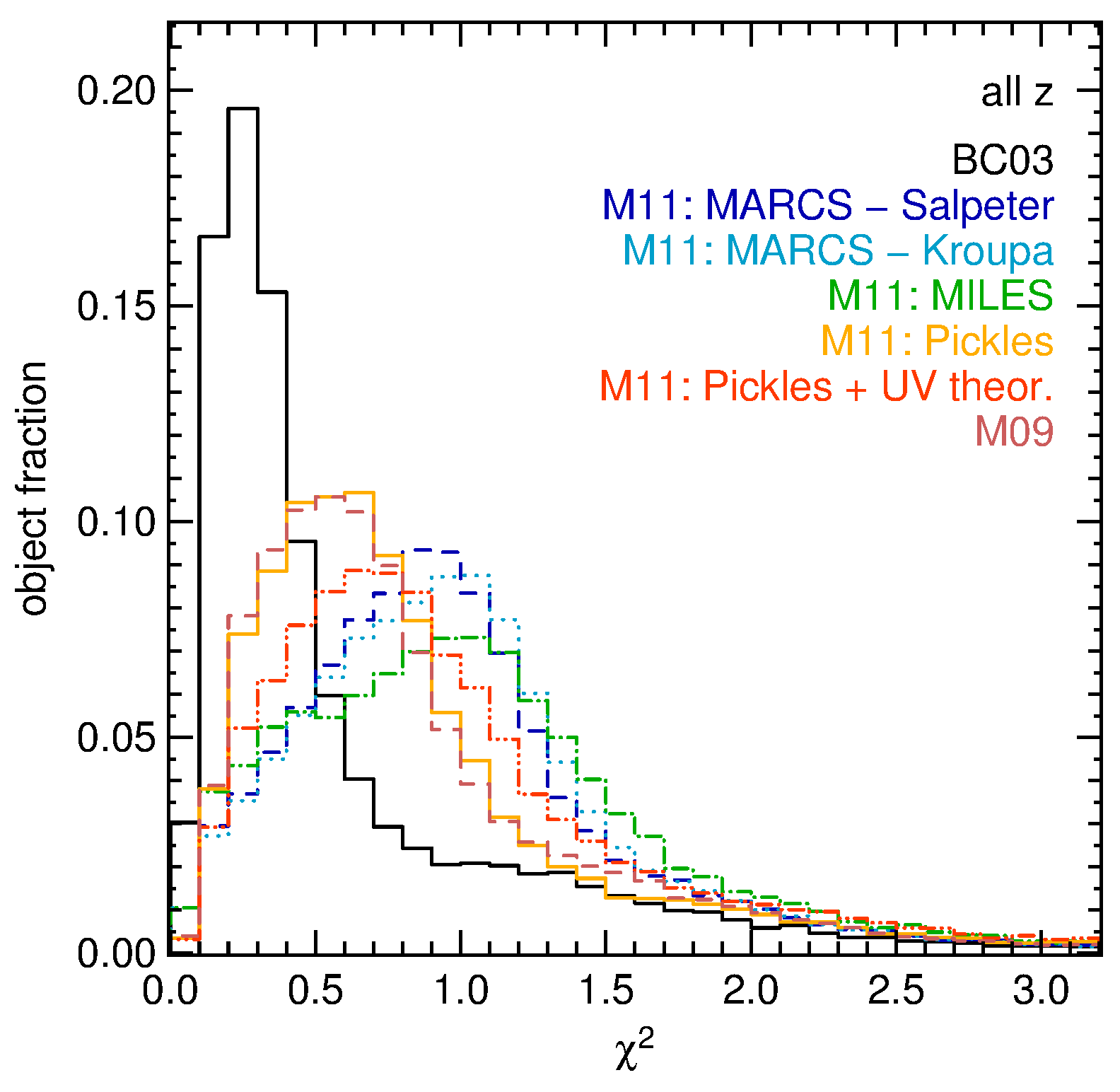}
\caption{
$\chi^2$ distribution of \sedfit\ results of different model-sets:
BC03 (black), M11 MARCS stellar library with Salpeter IMF (blue), MARCS with Kroupa IMF (light blue), M11 MILES (green), M11 Pickles (yellow), Pickles with theoretical UV (red), and M09 (light red).
The different panels show the various redshift slices in which the fitting was performed.
The lower right panel finally displays the $\chi^2$ results in all redshift intervals combined.
}
\label{fig:SEDfitchi2}
\end{figure}

In Section~\ref{sec:sedfitmodels} we created templates by fitting model SEDs to the photometry of the SDSS LRGs in order to produce SEDs matching the colors of the data.
Here we want to show the fitting results and how well the models are able to cover the observed colors.
In Figures~\ref{fig:ZvsCOL_bc03} and \ref{fig:ZvsCOL_M09} we saw that both the BC03 and the M09 model SEDs can represent the LRG data, at least for a large part of the investigated redshift range.
Both sets have a large variety of ages (plus $e$-folding timescales and metallicities in the case of BC03 models), thus they offer the perspective to provide a well matching model for each redshift.\\
In addition to that, \citet[][M11]{2011MNRAS.418.2785M} provide a large variety of single stellar populations (SSPs) created from different stellar libraries and with varying metallicities and ages.
The M11 models are promising candidates to represent the LRG data in terms of colors, wherefore we analyzed them in this regard and compared to the results with BC03 models.
We could only use those M11 model SEDs that are provided for a wavelength range sufficiently large to cover all SDSS filters at all LRG redshifts.
These are two flux-calibrated empirical stellar spectral libraries, Pickles \citep{1998PASP..110..863P} and MILES \citep{2006MNRAS.371..703S}, and the theoretically derived MARCS \citep{2008A&A...486..951G} library.
The Pickles models have solar metallicity, either an IMF of  \cite{2003PASP..115..763C} or Salpeter, and were extended in the near-IR by M11.
A version with a theoretical UV part was additionally created by M11.
The MILES SSPs in question were extended in the near-IR and UV, have a revised IR slope, and solar metallicity.
The MARCS models have metallicities of $Z=0.0004, 0.02$, and $0.04$, and IMFs of Salpeter and \citet{2001MNRAS.322..231K} respectively...
A more thorough description of the SSPs as well as the SEDs can be found in M11 and at \url{http://www.icg.port.ac.uk/\textasciitilde maraston/M11/}.
These models are a subset of all M11 covering a wavelength range wide enough for photometric redshift estimation.\\
We produced CSPs from the selected M11 models with the software \ezgal\footnote{\url{http://www.baryons.org/ezgal/}} \citep{2012PASP..124..606M}.
We apply exponential SFRs and the same $e$-folding timescales that were used with BC03 (Section~\ref{sec:BC03colors}).
The whole range of available ages, IMFs, and metallicities were exploited where possible in order to cover the greatest variety within the resulting SEDs.\\
These CSPs were then fitted to the photometry in the same way as the BC03 CSPs in Section~\ref{sec:sedfitmodels} (i.e., superposed by a burst model and with additional extinction) for each stellar library and IMF separately.
The same was also done with M09 models.\\
The SED fitting results for the different redshift bins and color spaces of M09/11 are shown together with those of the BC03 models in Figure~\ref{fig:COLSEDfit}.
From the M09/11 models in all cases the one with the smallest $\chi^2$ result is chosen for the plot.
The upper left panel of Figure~\ref{fig:COLSEDfit} ($z\approx0.02$) shows that the BC03 models cover the LRG colors better in this $z$ range, with the exception of $riiz$, where M09/M11 account for the lower right part that is not covered by BC03.
Furthermore, the BC03 color contours trace the LRG distribution better than M09/11 in $grri$ and $griz$.
In the upper right and middle left panel the colors are shown for the redshift bins at $z\approx0.1$, and $z\approx0.2$.
Both model sets perform similarly well, except for $z\approx0.2$, where M09/11 produces a slightly wider spread in $ugri$.
Apart from that, the BC03 contours follow well the density of the LRGs in the color spaces $grri$ and $griz$.
At $z\approx0.3$ the M09/11 models outperform BC03 in $riiz$, whereas both describe the data colors similarly in the other spaces (middle right panel).
Here, the M09/11 SEDs produce two peaks in $riiz$ which cannot be observed for the LRGs.
Finally, the lower panel of Figure~\ref{fig:COLSEDfit} shows the redshift bin around $0.4$, where M09/M11 still produce a wider spread in $riiz$, but are not able to match the bluer part of the LRG's $r-i$ color.
In addition to that, neither BC03 nor M09/11 models cover the bluer part of the $i-z$ color.
In Section~\ref{sec:sedfitmodelselection} we already mentioned that the spread in the colors of the data cannot be completely accounted for by the photometric errors in this redshift range.
The reason for the spread must thus be of physical origin.
Unfortunately, not the whole range of $i-z$ of LRGs at $z\approx0.4$ can be fitted by the investigated models sets while simultaneously matching the $g-r$ and $r-i$ colors (where the latter has smaller errors than $i-z$).
This is a feature of BC03 and M09/11 models, and the reason why we could not create SEDs with bluer $i-z$ in Section~\ref{sec:sedfitmodels}.\\
The distributions of the $\chi^2$ results of the fitting to the various model sets are shown in Figure~\ref{fig:SEDfitchi2}.
Each analyzed $z$ bin is displayed in a different panel, whereas the $\chi^2$ distributions for all redshift intervals combined is shown in the lower right panel.
BC03 outperforms the other models (M09 and M11), regarding the $\chi^2$ values, in every redshift slice except for $z\approx0.4$, where the M11 Pickles models fit the LRGs better, but only marginally.
We would like to add that the M11 Pickles models fit the data best compared to the other M09/M11 SEDs, and outperform the Pickles models with a theoretical UV-part of the spectrum for each fitted object.\\
Finally, the lower right panel of Figure~\ref{fig:SEDfitchi2} shows that the $\chi^2$ values are best also for all $z$ bins combined, as expected from the previous results.
Following these results, and because of the previous analysis of the colors of the produced SEDs, we adopted the BC03 models for the creation of photo-$z$ template SEDs.

\section{Color\textendash Color Relations of the New Models}
\label{app:colcol}
\begin{figure}
\centering
\includegraphics[width={0.55\textwidth}]{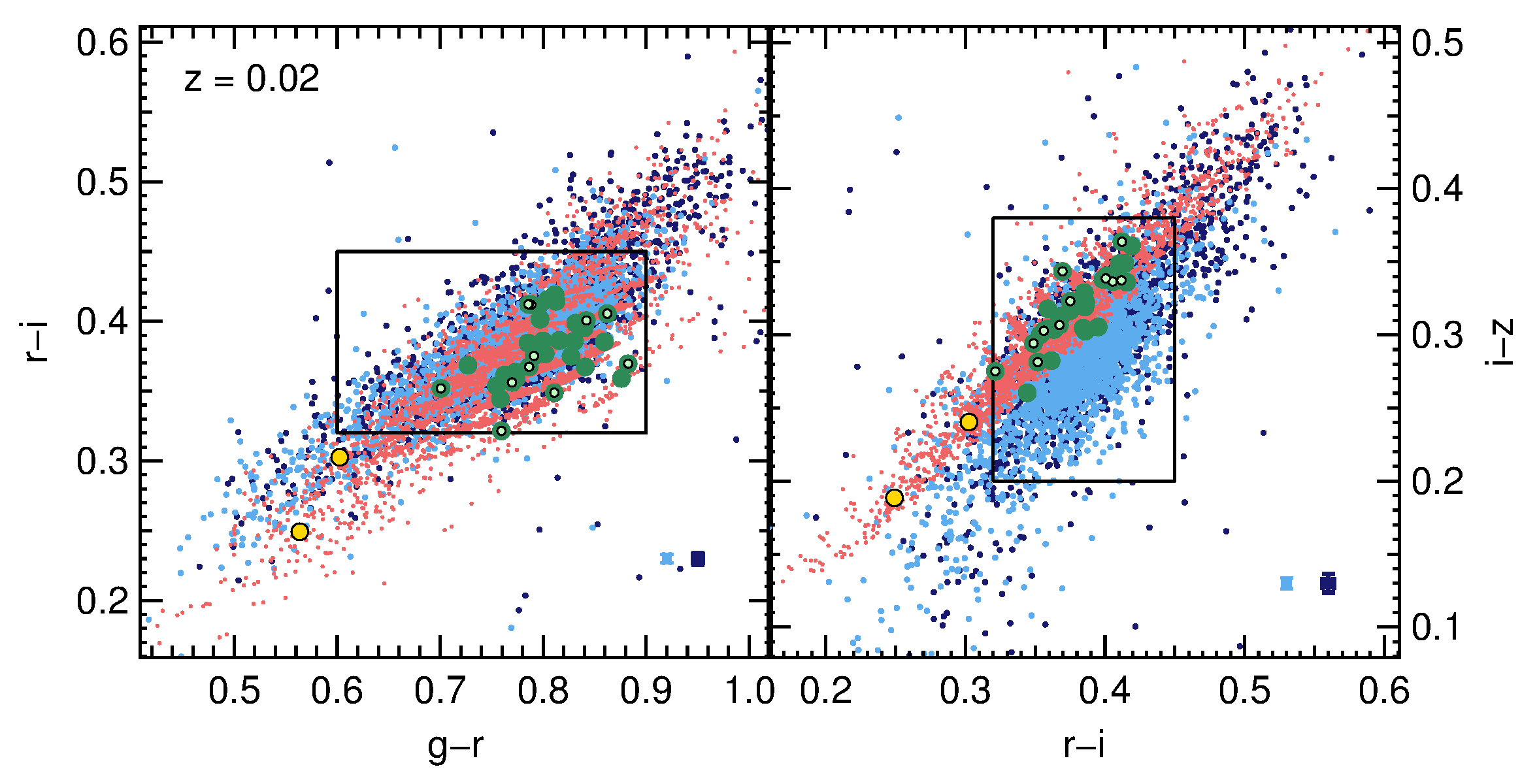}
\caption{Color vs. color plots for SDSS data (blue) and their individual best fitting \sedfit-SEDs (red) for $0.0\leq z\leq0.04$.
	  Objects with $u$ band errors lower than the median are indicated in light blue, whereas those with greater errors are dark blue.
	  The mean error bars of all objects are shown in the lower right corners in dark blue, and the mean errors of objects with $u$ band errors smaller than the median in light blue.
	  The grid in $r-i$ and the boundaries within which the models are selected are shown in black.
	  The dark green dots are the preselected models, whereas the light green points represent the models that are left over after the removal of redundant SEDs.
	  The yellow dots are models that shall account for objects outside the selected boundaries in $ugri$ and $griz$.}
\label{fig:selectbycol_z00-2}
\end{figure}

\begin{figure}
\centering
\includegraphics[width={0.55\textwidth}]{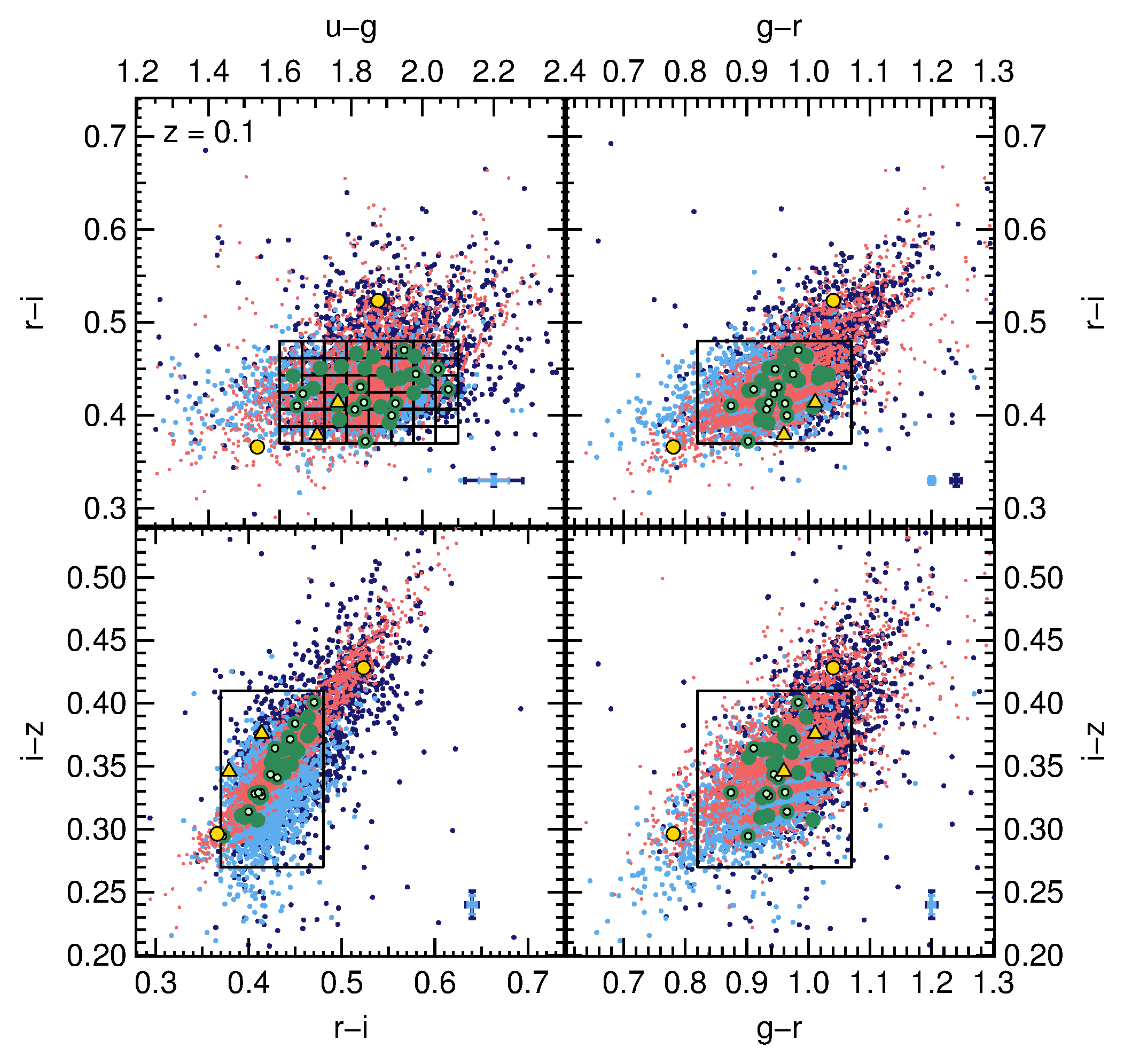}
\caption{
Color vs. color plots for SDSS data (blue) and their individual best fitting \sedfit-SEDs (red) for $0.08\leq z\leq0.12$.
	  Objects with $u$ band errors lower than the median are indicated in light blue, whereas those with greater errors are dark blue.
	  The mean error bars of all objects are shown in the lower right corners in dark blue, and the mean errors of objects with $u$ band errors smaller than the median in light blue.
	  The grid (upper panel) and the boundary (lower panel) within which the models are selected are shown in black.
	  The dark green dots are the preselected models, whereas the light green points represent the models that are left over after the removal of redundant SEDs.
	  The yellow dots are models that shall account for objects outside the selected boundaries in $ugri$ and $griz$.
	  Yellow triangles are models from $z\approx0.02$ that lie in the lower part of the UV red sequence of R12 (see Section~\ref{sec:UV}, Figure~\ref{fig:UV}) and are added additionally in this redshift range.
}
\label{fig:selectbycol_z01}
\end{figure}

\begin{figure}
\centering
\includegraphics[width={0.55\textwidth}]{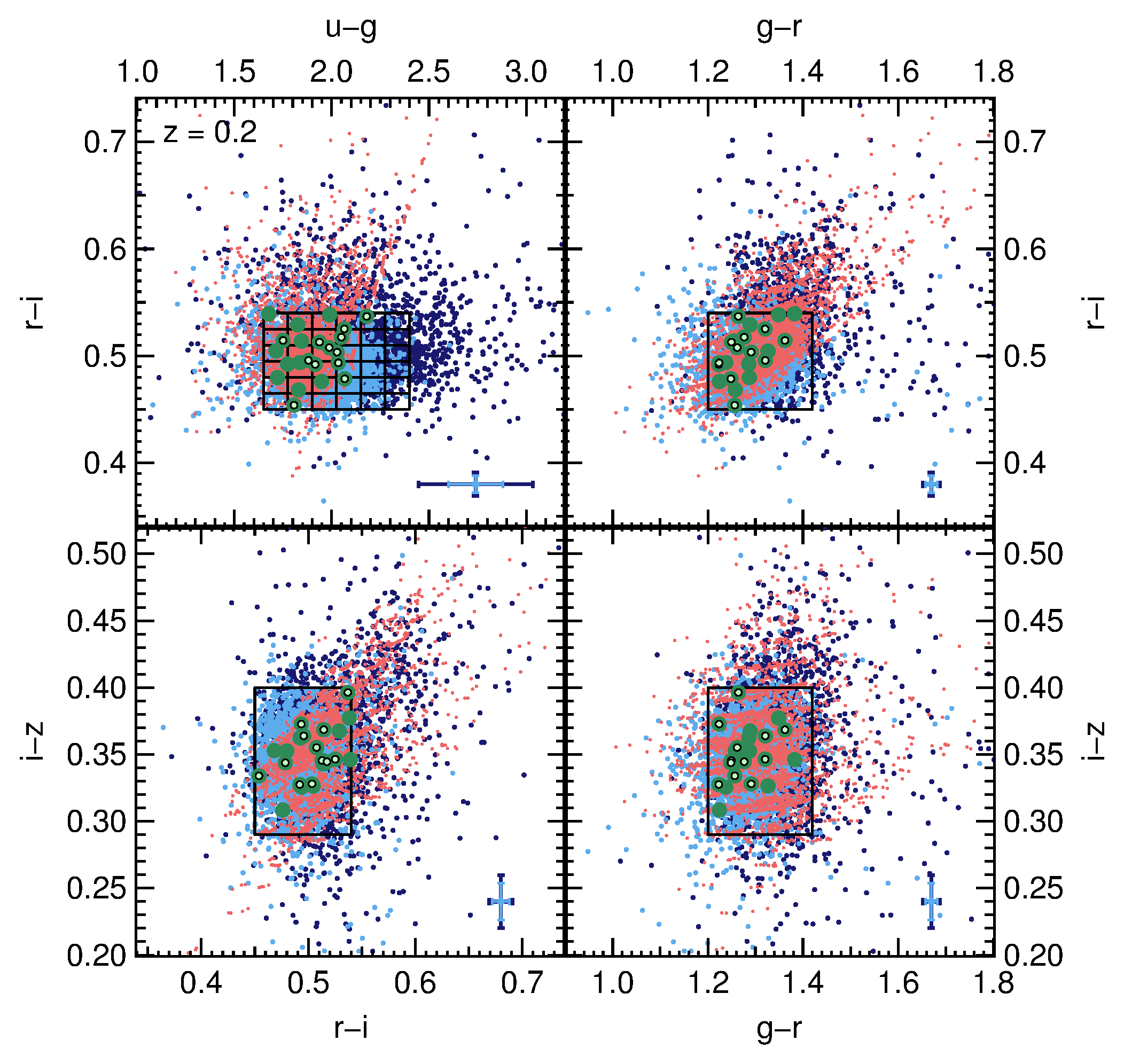}
\caption{
Color vs. color plots for SDSS data (blue) and their individual best fitting \sedfit-SEDs (red) for $0.18\leq z\leq0.22$.
	  Objects with $u$ band errors lower than the median are indicated in light blue, whereas those with greater errors are dark blue.
	  The mean error bars of all objects are shown in the lower right corners in dark blue, and the mean errors of objects with $u$ band errors smaller than the median in light blue.
	  The grid (upper panel) and the boundary (lower panel) within which the models are selected are shown in black.
	  The dark green dots are the preselected models, whereas the light green points represent the models that are left over after the removal of redundant SEDs.
}
\label{fig:selectbycol_z02}
\end{figure}

\begin{figure}
\centering
\includegraphics[width={0.55\textwidth}]{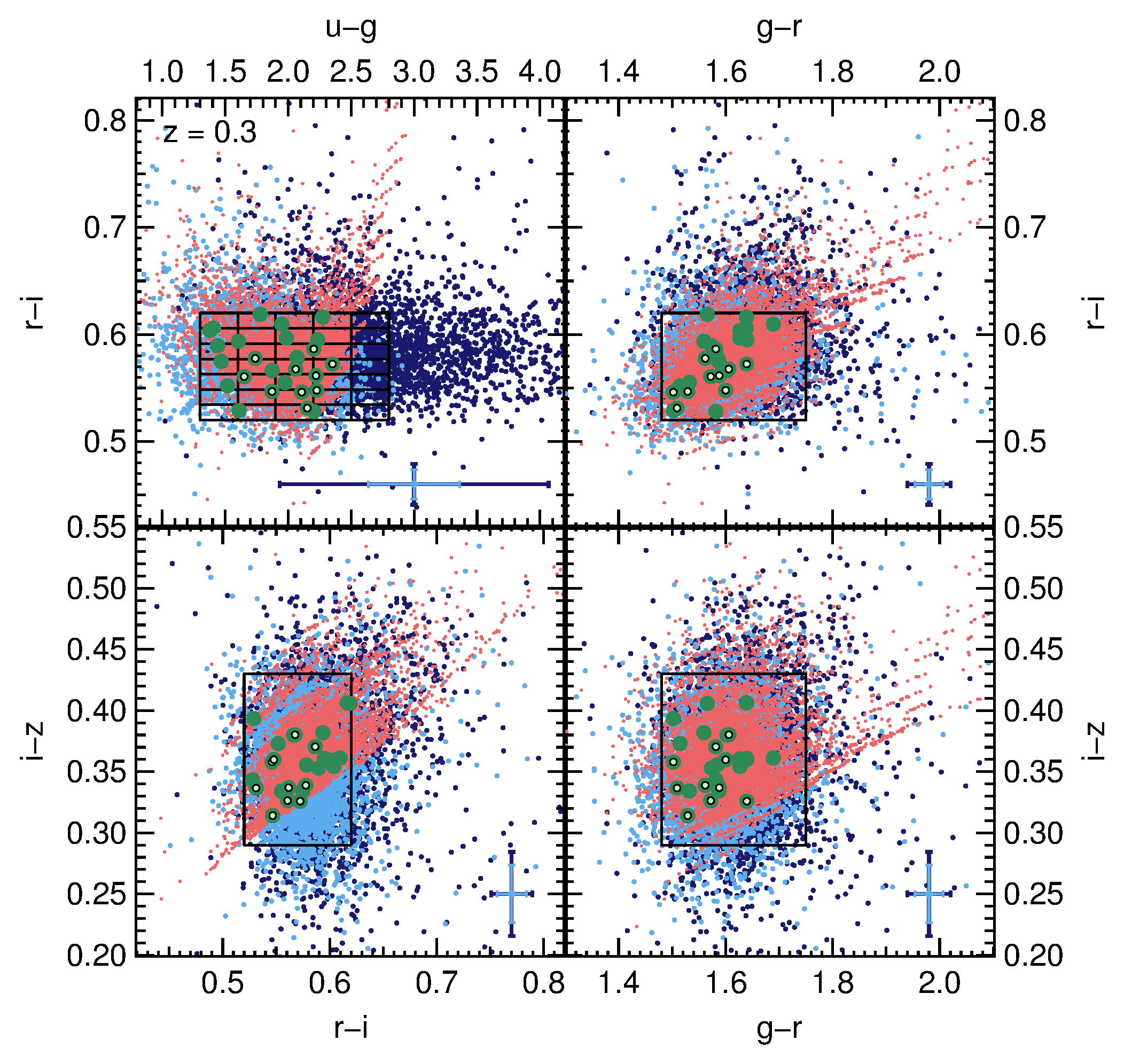}
\caption{
Color vs. color plots for SDSS data (blue) and their individual best fitting \sedfit-SEDs (red) for $0.28\leq z\leq0.32$.
	  Objects with $u$ band errors lower than the median are indicated in light blue, whereas those with greater errors are dark blue.
	  The mean error bars of all objects are shown in the lower right corners in dark blue, and the mean errors of objects with $u$ band errors smaller than the median in light blue.
	  The grid (upper panel) and the boundary (lower panel) within which the models are selected are shown in black.
	  The dark green dots are the preselected models, whereas the light green points represent the models that are left over after the removal of redundant SEDs.
}
\label{fig:selectbycol_z03}
\end{figure}

\begin{figure}
\centering
\includegraphics[width={0.55\textwidth}]{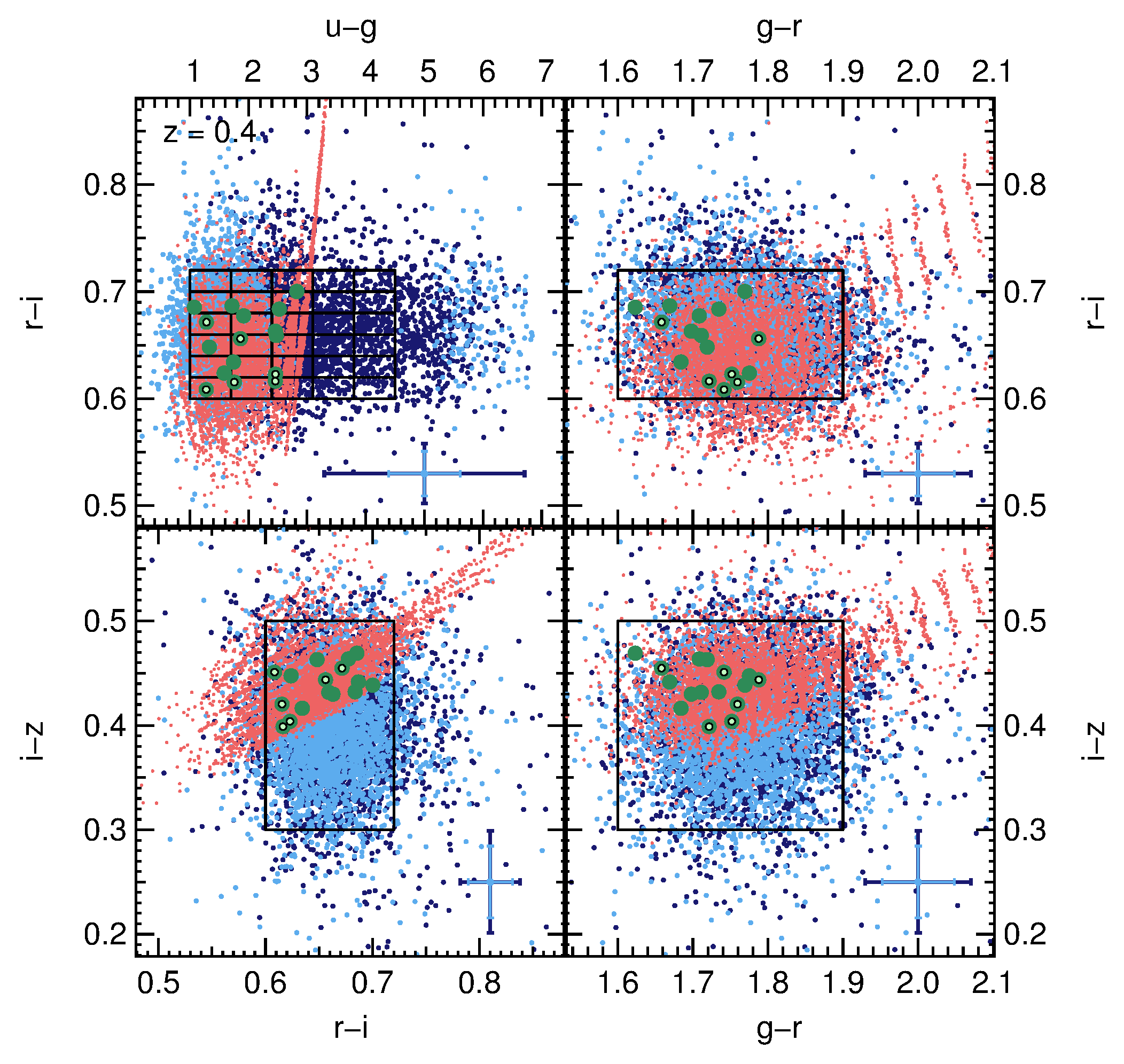}
\caption{
Color vs. color plots for SDSS data (blue) and their individual best fitting \sedfit-SEDs (red) for $0.38\leq z\leq0.42$.
	  Objects with $u$ band errors lower than the median are indicated in light blue, whereas those with greater errors are dark blue.
	  The mean error bars of all objects are shown in the lower right corners in dark blue, and the mean errors of objects with $u$ band errors smaller than the median in light blue.
	  The grid (upper panel) and the boundary (lower panel) within which the models are selected are shown in black.
	  The dark green dots are the preselected models, whereas the light green points represent the models that are left over after the removal of redundant SEDs.
}
\label{fig:selectbycol_z04}
\end{figure}

Here we present the color\textendash color relations of our models and the LRG data.
Figures~\ref{fig:selectbycol_z00}, and \ref{fig:selectbycol_z00-2}\textendash\ref{fig:selectbycol_z04} show the colors of the best fitting SED models, the preselected models, and the selected models after the removal of redundant SEDs in comparison to the LRG data itself in the $ugri$, $griz$, $grri$, and $riiz$ color spaces.
We display the LRGs with $u$ band errors below the median and those with errors above with different colors to visualize the difference in the populated loci within the $ugri$ space, which becomes evident at $z\approx0.3$ (Figure~\ref{fig:selectbycol_z03}, upper panel) and $z\approx0.4$ (Figure~\ref{fig:selectbycol_z04}, upper panel).\\
The grids we imposed on the $ugri$ plane within which we selected templates (see Section~\ref{sec:sedfitmodelselection}) are displayed in the upper left panels of Figures~\ref{fig:selectbycol_z00} , and \ref{fig:selectbycol_z01}\textendash\ref{fig:selectbycol_z04}, whereas the color boundaries in $g-r$ and $i-z$ are drawn in the lower panel of Figure~\ref{fig:selectbycol_z00}, in Figure~\ref{fig:selectbycol_z00-2} and in the lower left and in the right panels of Figures~\ref{fig:selectbycol_z01}\textendash\ref{fig:selectbycol_z04}.\\
In order to account for objects with peculiar colors which lie outside the boundaries as well, we select additional SEDs by eye and confirm that they yield a photometric redshift bias lower than 0.01, when used as single template for the subcatalog from the same $z$ range.
If they do not meet that requirement, we chose the SED of another object from the color neighborhood of the first one, test for the bias, and so on until we find a model from that color locus which creates only a small $\zphot$-bias.\\
Bright cluster galaxies (BCGs) with very low UV-excess (i.e., in the lower part of the UV red sequence; see Section~\ref{sec:UV}, Figure~\ref{fig:UV}) at $z\sim0.1$ are not matched by the models already selected for that $z$ range.
But we find that two SEDs originating from $z\approx0.2$ are able to cover the considered region.
Therefore, we add these templates to $z\approx0.1$, i.e., we assign the same redshift priors as for $z\approx0.1$ models in addition to their former priors (see Section~\ref{sec:PhotoZ}), thereby covering the lower UV red sequence.
The colors of these two models are shown by yellow triangles in Figure~\ref{fig:selectbycol_z01}.

\end{document}